\documentclass[fleqn,10pt]{wlscirep}
\usepackage[utf8]{inputenc}
\usepackage[T1]{fontenc}

\usepackage{todonotes}
\usepackage{color}
\usepackage{subfig}
\usepackage{booktabs}
\usepackage{flushend}
\usepackage[export]{adjustbox}
\usepackage{xr}

\makeatletter
\newcommand*{\addFileDependency}[1]{
  \typeout{(#1)}
  \@addtofilelist{#1}
  \IfFileExists{#1}{}{\typeout{No file #1.}}
}
\makeatother

\newcommand*{\myexternaldocument}[1]{%
    \externaldocument{#1}%
    \addFileDependency{#1.tex}%
    \addFileDependency{#1.aux}%
}

\myexternaldocument{./supplementary}

\title{Artificial intelligence for online characterization of ultrashort X-ray free-electron laser pulses}

\author[a,f,1,*]{Kristina Dingel}
\author[a,1]{Thorsten Otto}
\author[b,f]{Lutz Marder}
\author[e]{Lars Funke}
\author[e]{Arne Held}
\author[e]{Sara Savio}
\author[b,f]{Andreas Hans}
\author[c,f]{Gregor Hartmann}
\author[a,c,f]{David Meier}
\author[c,f]{Jens Viefhaus}
\author[a,f]{Bernhard Sick}
\author[b,f]{Arno Ehresmann}
\author[g,d,b]{Markus Ilchen}
\author[e]{Wolfram Helml}

\affil[*]{kristina.dingel@uni-kassel.de}

\affil[a]{Intelligent Embedded Systems, University of Kassel, Wilhelmsh\"{o}her Allee 73, 34121 Kassel, Germany}
\affil[b]{Institute for Physics and CINSaT, University of Kassel, Heinrich-Plett-Stra\ss e~40, 34132 Kassel, Germany}
\affil[c]{Helmholtz-Zentrum Berlin f\"{u}r Materialien und Energie, Hahn-Meitner-Platz 1, 14109 Berlin, Germany}
\affil[d]{European XFEL GmbH, Holzkoppel 4, 22869 Schenefeld, Germany}
\affil[e]{Technische Universität Dortmund, Fakultät Physik, Maria-Goeppert-Mayer-Straße 2, 44227 Dortmund, Germany.}
\affil[f]{Artificial Intelligence Methods for Experiment Design (AIM-ED), Joint Lab Helmholtzzentrum für Materialien und Energie, Berlin (HZB) and University of Kassel
}
\affil[g]{Deutsches Elektronen-Synchrotron DESY, Notkestr. 85, 22607 Hamburg, Germany}

\affil[1]{these authors contributed equally to this work}

\keywords{Angular streaking, Detector image analysis, Free-electron laser, Pulse characterization, Machine learning, Artificial intelligence, Simulation, Convolutional neural network}

\begin{abstract}

X-ray free-electron lasers (XFELs) as the world's brightest light sources provide ultrashort X-ray pulses with a duration typically in the order of femtoseconds. Recently, they have approached and entered the attosecond regime, which holds new promises for single-molecule imaging and studying nonlinear and ultrafast phenomena such as localized electron dynamics. The technological evolution of XFELs toward well-controllable light sources for precise metrology of ultrafast processes has been, however, hampered by the diagnostic capabilities for characterizing X-ray pulses at the attosecond frontier. In this regard, the spectroscopic technique of photoelectron angular streaking has successfully proven how to non-destructively retrieve the exact time--energy structure of XFEL pulses on a single-shot basis. By using artificial intelligence techniques, in particular convolutional neural networks, we here show how this technique can be leveraged from its proof-of-principle stage toward routine diagnostics even at high-repetition-rate XFELs, thus enhancing and refining their scientific accessibility in all related disciplines.

\end{abstract}

\begin{document}

\flushbottom
\maketitle

\thispagestyle{empty}

\section*{\label{sec:Introduction}Introduction}

X-ray free-electron lasers (XFELs) are the world's fastest X-ray cameras, providing ultrashort exposure times in combination with a spatial resolution limit down to the sub-nanometer range, which allows for time-resolved experiments \textit{'freezing'} the motion of atoms and molecules. In fact, XFELs have revolutionized several fields of science enabling us to observe the role of transient structures and resonances in atoms \cite{mazza2020mapping} as well as single-molecule or cluster imaging \cite{SingleMolImTrans}, investigations of ultrafast processes at element-specific observer sites \cite{Roadmap}, and the study of nonlinear light--matter interaction in the X-ray regime \cite{eichmann2020photon}.

Over the past decade, further development of the underlying machine operation techniques has enabled increasingly sophisticated control over the photon pulse parameters. One of the most recent major upgrades is the increased repetition rate of XFELs that is anticipated to initiate a leap from proof-of-principle experiments to advanced applications of interdisciplinary importance, thus representing a cornerstone of modern XFEL science \cite{decking2020mhz}.

Most of the FELs and in fact all XFELs worldwide are currently based on the principle of \textit{self-amplification of spontaneous emission} (SASE) \cite{milton2001exponential}. More precisely, their pulses are formed stochastically through the interplay between the relativistically accelerated electron bunches themselves and the spontaneously emitted synchrotron radiation, caused by their sinusoidal trajectories inside magnetic structures with periodically changing polarity, so-called \textit{undulators}. This feedback interaction leads to a subsequent density modulation of the undulating electrons, resulting in bursts of ultrashort X-ray pulses with a peak brightness up to and exceeding $10^{32}$ $\rm \frac{photons}{sec \cdot mrad^2 \cdot mm^2 \cdot 0.1\% BW}$ (with bandwidth $BW$). The amplification process generates a non-predictable time--energy structure for every single pulse, constituting one of the biggest limitations for XFEL science so far. There are currently no control mechanisms and no routinely available diagnostics in place to \textit{directly} measure the temporal properties of these X-ray pulses. This is due to the stochastic nature of each individual XFEL pulse, making their single-shot characterization necessary and precluding standard integrated methods developed for attosecond pulses from lab-based sources \cite{Tzallas2003, Paul2001, Kienberger2004}. Hence, the bulk of dynamics on attosecond to femtosecond time scales occurring during the exposure to the X-rays is unfortunately only inferred via indirect pulse measurements such as spectral analysis \cite{serkez2020opportunities} or electron beam diagnostics \cite{behrens2014few}.

Recently, we have demonstrated a new technique termed \textit{angular streaking} that is capable of retrieving the time--energy structure of all incoming SASE X-ray pulses non-destructively with attosecond resolution \cite{hartmann2018attosecond}. Besides the major diagnostic breakthrough, this generally paves the way for time-resolved and nonlinear attoscience in the X-ray regime. In fact, the onset of all structural dynamics in matter can now be studied in detail even from specific observer sites through strongly localized electrons. 
Fast and reliable feedback of this novel diagnostic regarding the experiment and the machine itself is of utmost importance for upcoming scientific applications at XFELs 
\cite{duris2020tunable, driver2020attosecond}. For high-repetition-rate XFELs such as the European XFEL near Hamburg, Germany, conventional analysis approaches fail to accommodate the enormous amount and complexity of angular-streaking data in full depth. Especially for online analysis and, ultimately, (re)active control and pulse shaping during beam times, conventional data processing methods are not suited.
Therefore, several core challenges of XFELs are anticipated to be tackled by artificial intelligence (AI), in particular machine-learning (ML) techniques.

\begin{figure}[!htbp]
	    \centering
	    \includegraphics[width=0.45\textwidth]{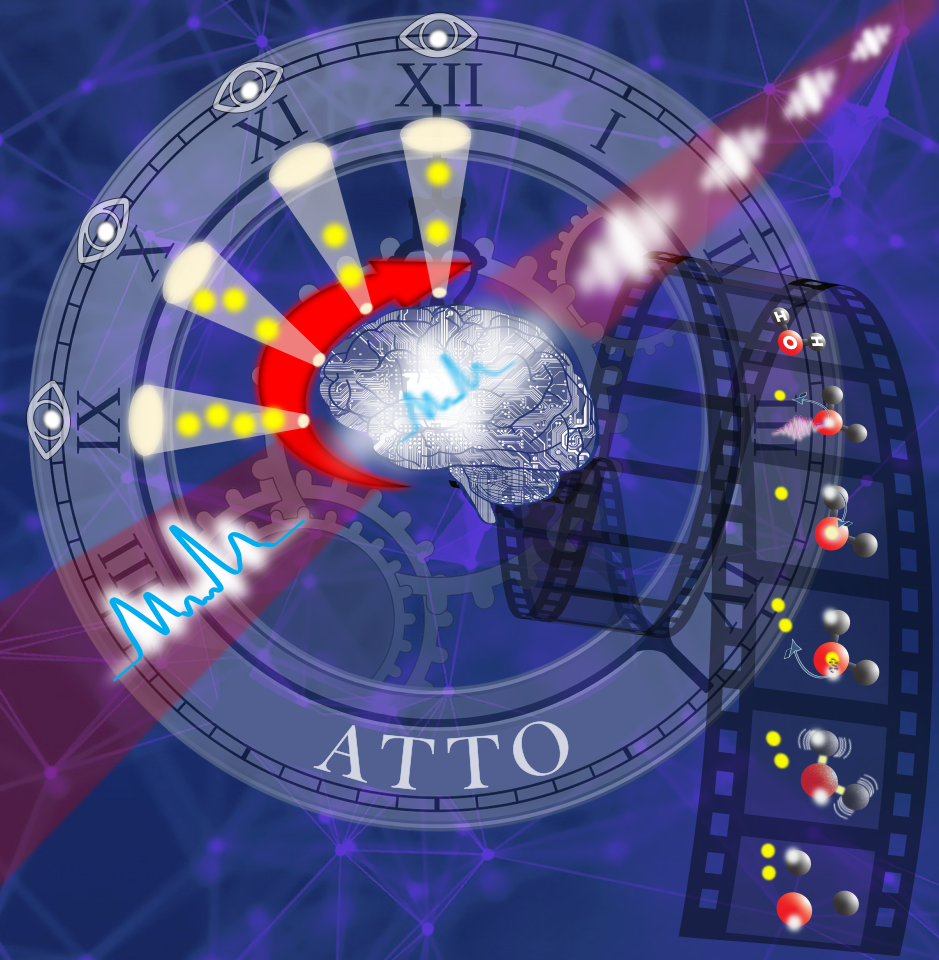}
	    \caption{Illustration of this work's topical orientation. A series of previously unknown XFEL pulses releases electrons (indicated in yellow) from a gas target with characteristic kinetic energies. Their kinetic energy shifts caused by a circularly polarized infrared streaking field are recorded in an angle-resolving spectrometer (indicated by the cones and eyes). In the artificial brain, the thus obtained information about the electrons is processed and reveals the underlying attosecond substructure of the X-ray pulses. The precise knowledge of the time–energy substructure can then be used to shape the X-ray pulses via a feedback loop to the XFEL machine settings or to enable access to ultrafast electron dynamics at the attosecond frontier, as here indicated by a molecular movie of an interatomic Auger decay in water with subsequent dissociation.}
	    \label{fig:ga}
	\end{figure}

In this article, we present a machine-learning-based proof-of-concept on retrieving the full and detailed XFEL pulse temporal profile, including the pulse duration and its intensity substructure. In addition, we show that it is possible to extract temporal information on the electronic processes after photoionization initiated by the X-rays, that is a subsequent Auger decay, via the method of angular streaking paired with analysis through neural networks (NNs). Moreover, by using simulated streaking data with various degrees of instrument noise and different electron emission signatures, we demonstrate the flexibility of the NN-based online diagnostic tool for XFELs. It is thus robust against detector noise and machine fluctuations, and covering the vast majority of current and future operation modes.

\begin{figure*}
    \centering
    \vspace{-5em}
    \subfloat[Exemplary shot showing a detector image measured during an angular streaking experiment. The image was taken for X-ray pulses at a photon energy of 1180 eV, ionizing photoelectrons from the 1s shell in neon, and streaked by a circularly polarized infrared laser with a wavelength of 10.6~$\mu$m (as published in \cite{hartmann2018attosecond}). ]{\includegraphics[width=0.49\textwidth]{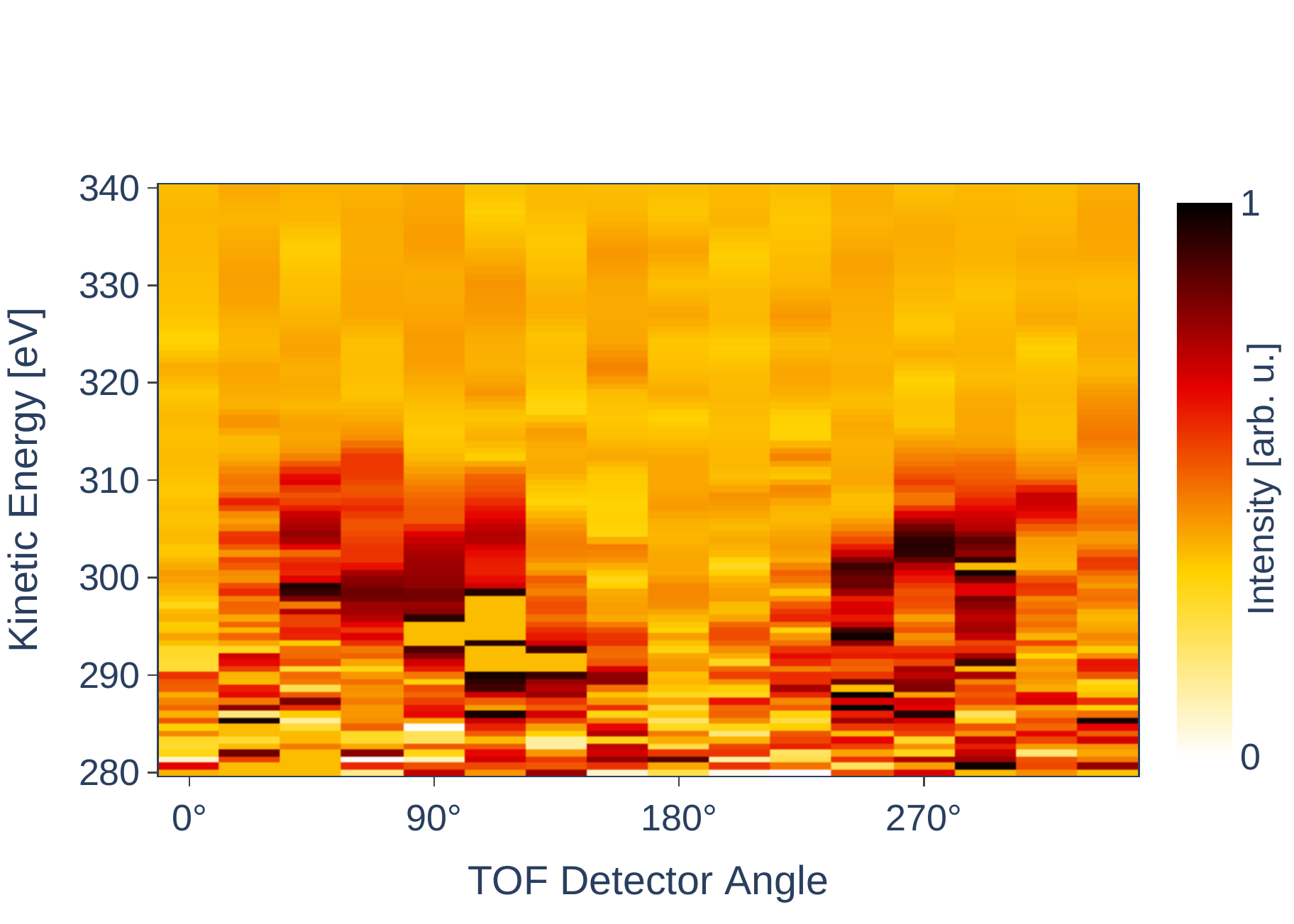}\label{fig:real_shot}}
    \hfill
    \subfloat[Example for a simulated detector image after adding $\pm30\ \%$ noise. The simulation parameters are chosen similar to those used in the experiment shown in panel (a).] {\includegraphics[width=0.49\textwidth]{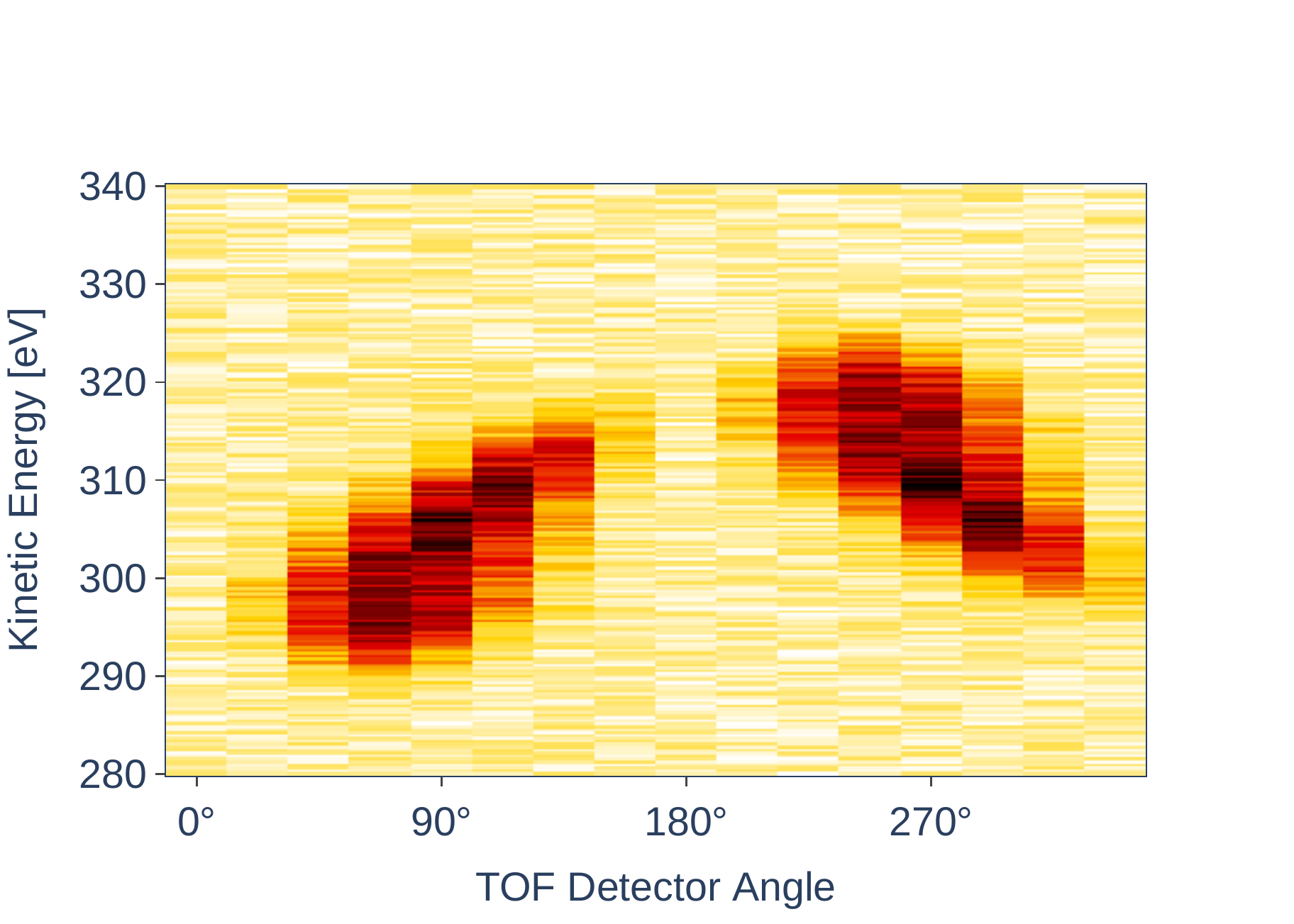}\label{fig:noiseLevels}}
    \vfill
    \subfloat[Simulated spectrogram of an X-ray pulse with a simple Gaussian intensity distribution.]{\includegraphics[width=0.33\textwidth]{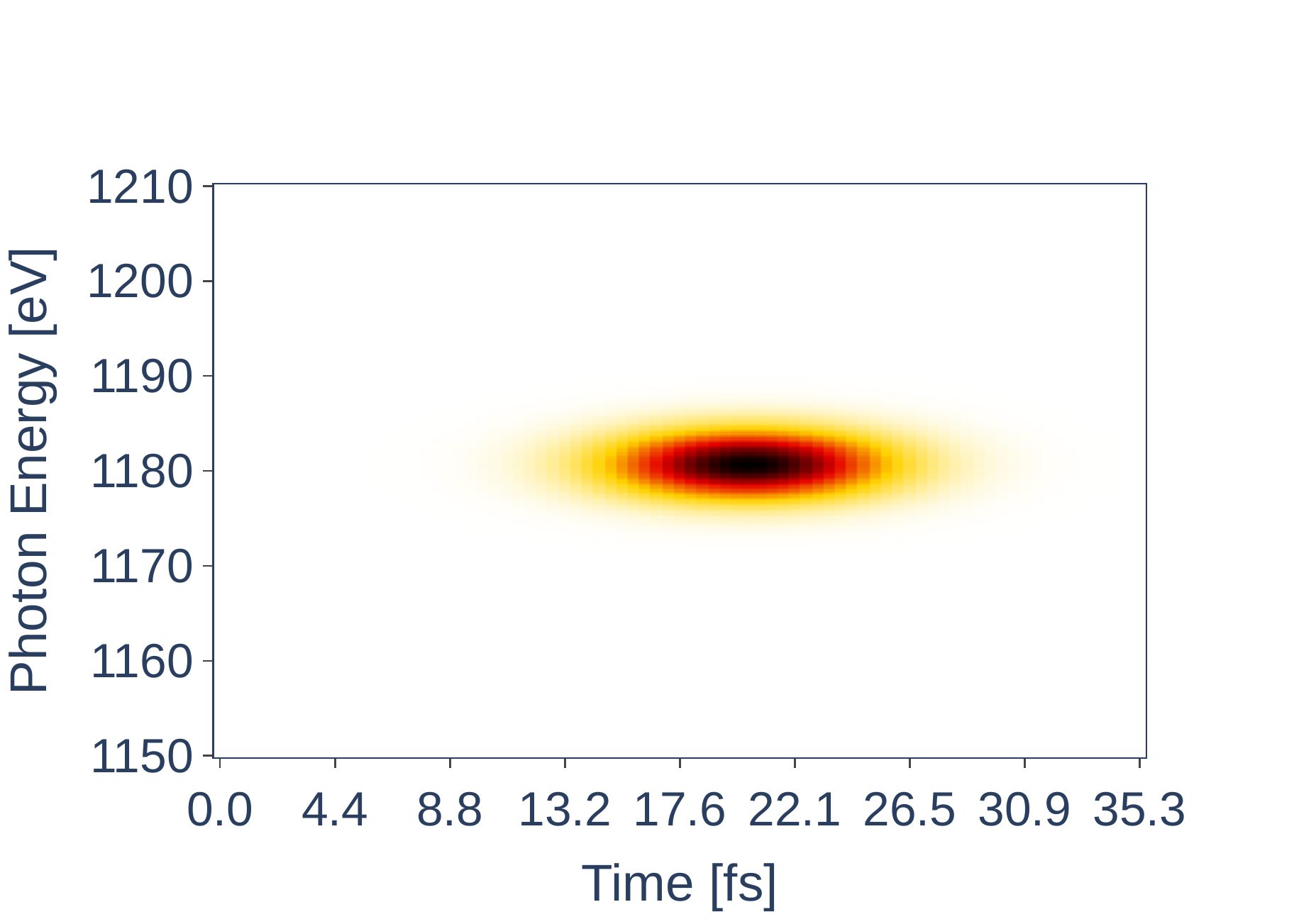}\label{fig:perfSpectrogram}}
    \hfill
    \subfloat[Simulated spectrogram after adding a spiky structure to mimick a SASE pulse.]{\includegraphics[width=0.33\textwidth]{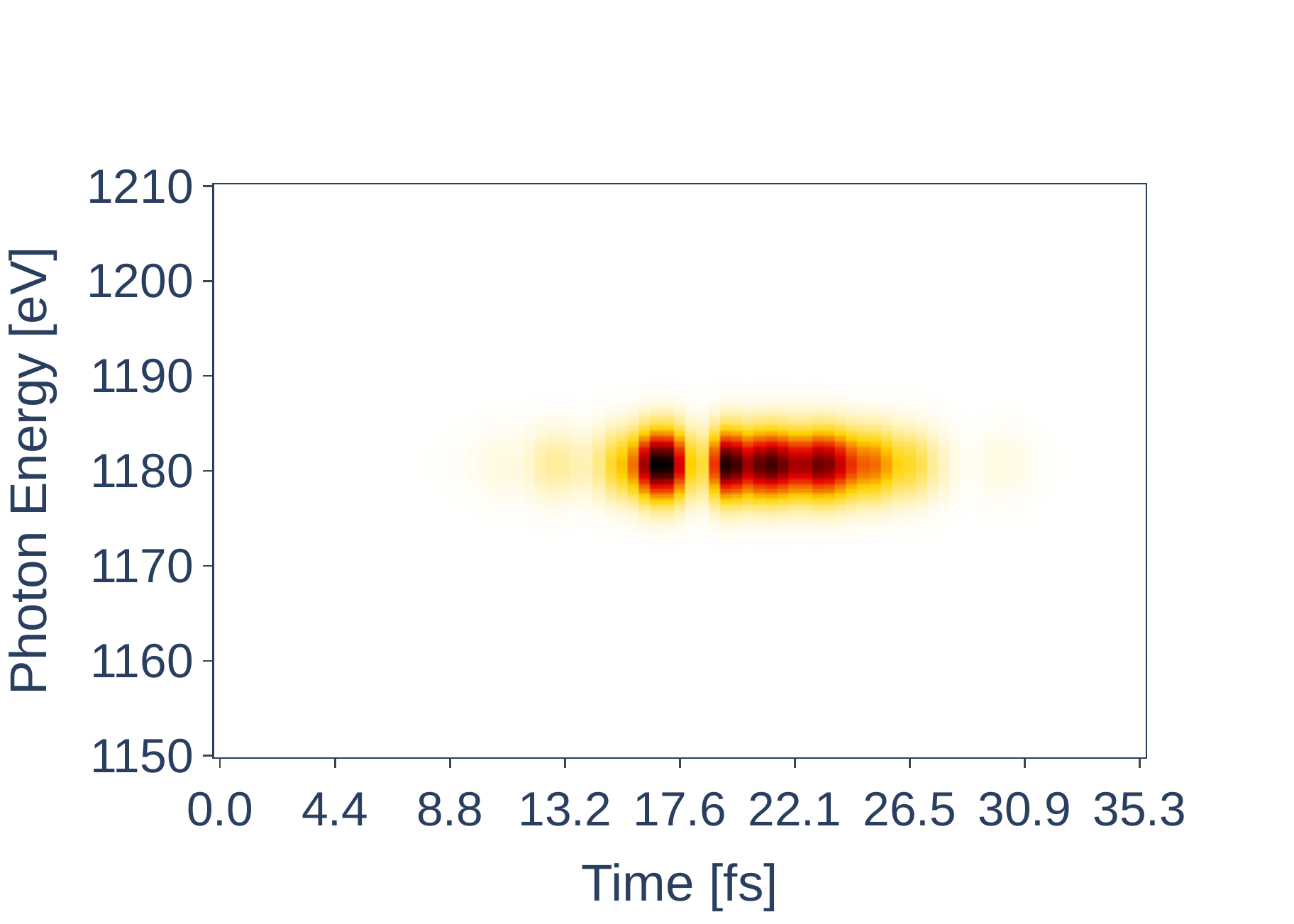}\label{fig:pulseSpectrogram}}
    \hfill
    \subfloat[Simulated spectrogram of Auger electrons after X-ray 1s ionization in neon for decay time of 7.6 fs.]{\includegraphics[width=0.33\textwidth]{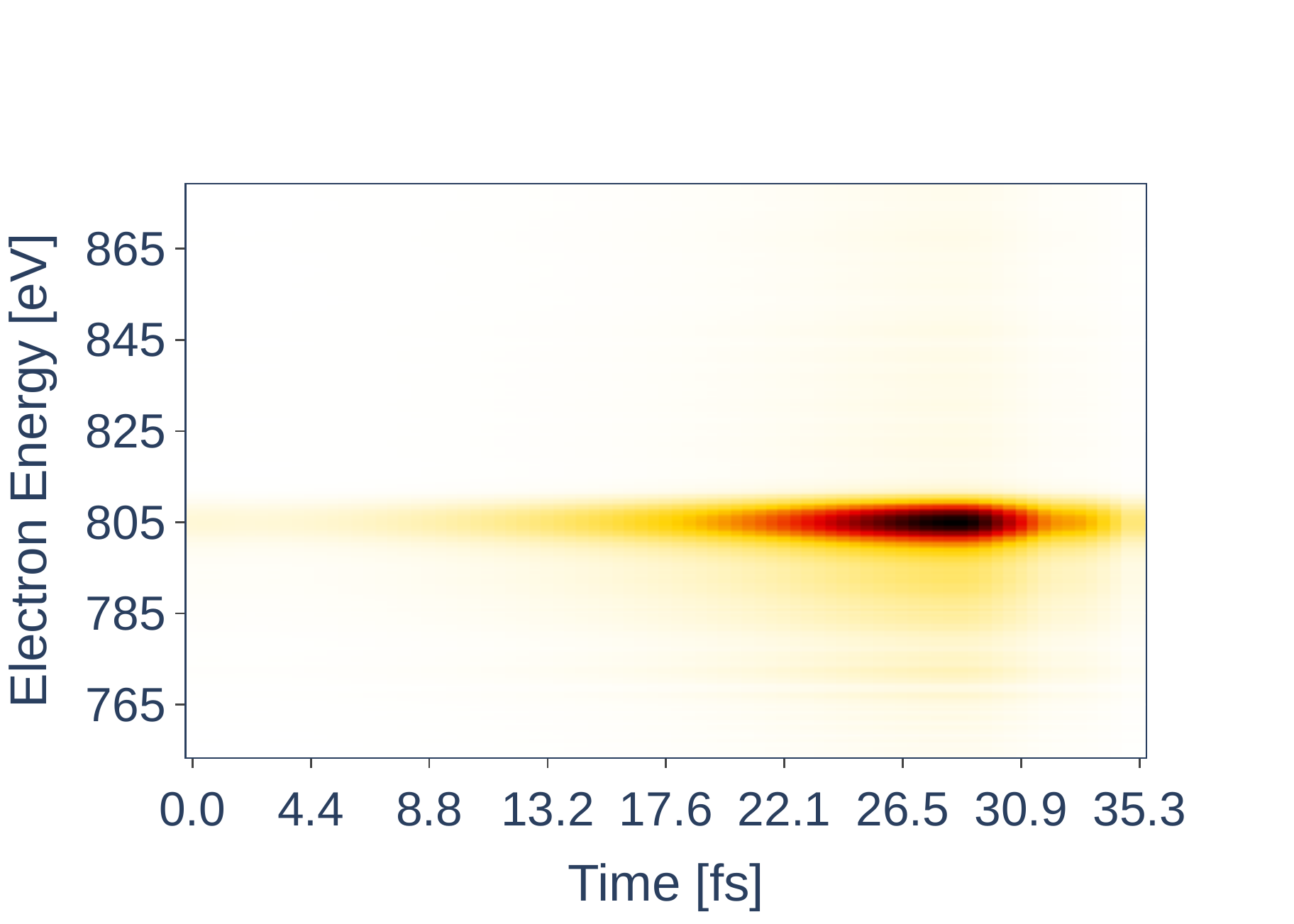}\label{fig:decaySpectrogram}}
    \vfill
    \subfloat[Simulated angularly streaked detector image generated from (c).]{\includegraphics[width=0.33\textwidth]{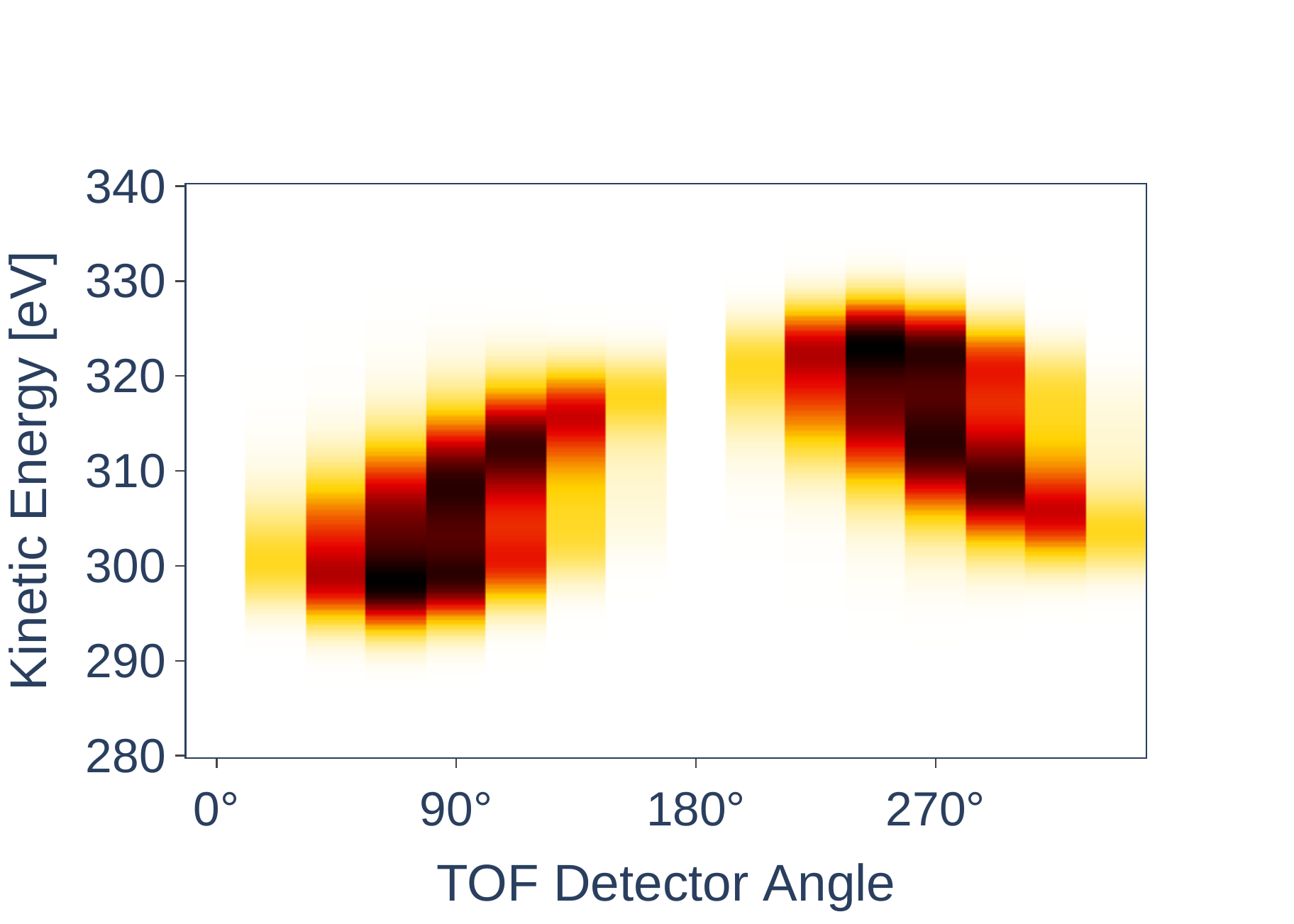}\label{fig:perfDetectorImage}}
    \hfill
    \subfloat[Simulated angularly streaked detector image generated from (d). Supplementary Fig.~1 highlights the difference between this detector image and (f).]{\includegraphics[width=0.33\textwidth]{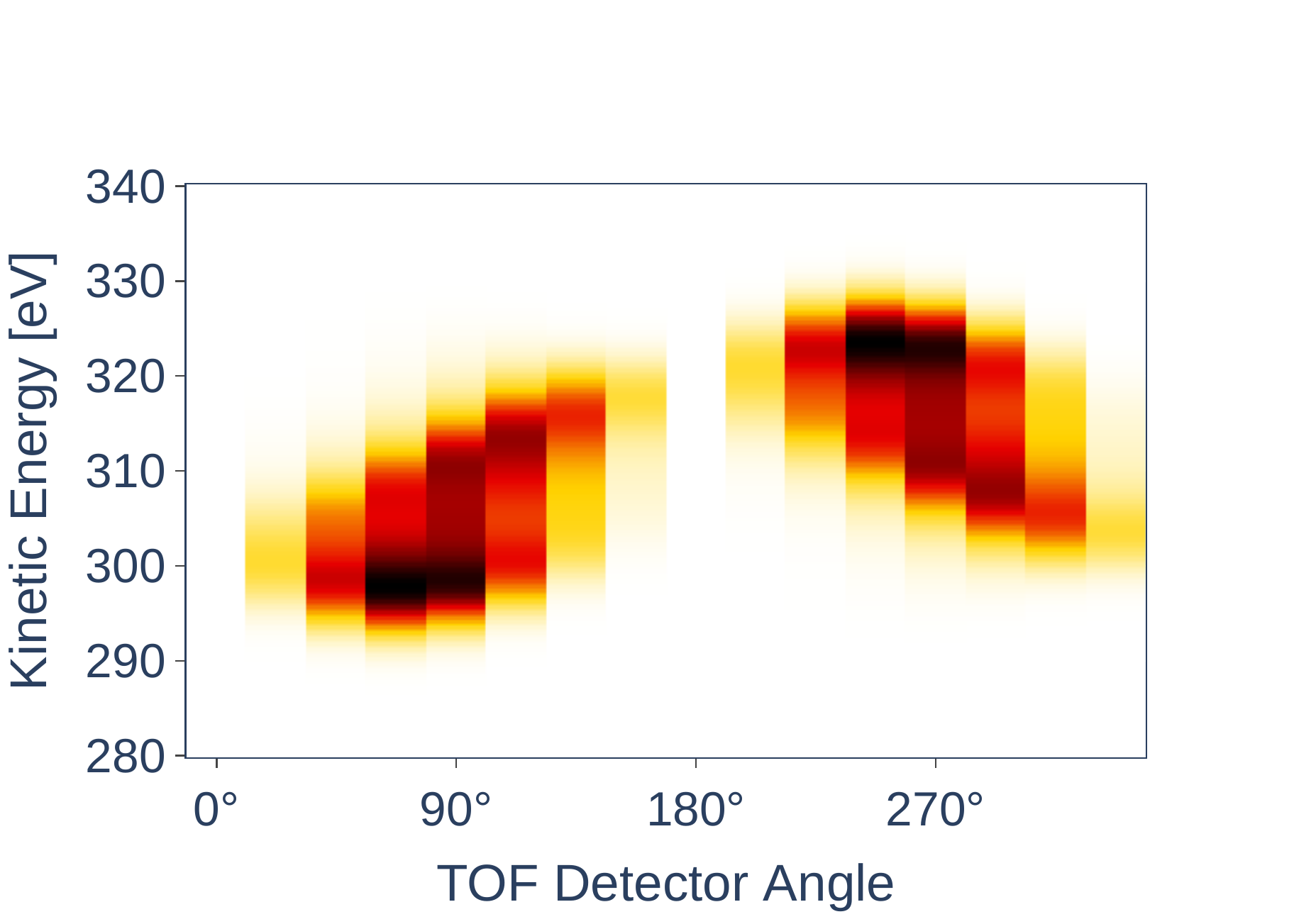}\label{fig:pulseDetectorImage}}
    \hfill
    \subfloat[Simulated angularly streaked detector image generated from (e).]{\includegraphics[width=0.33\textwidth]{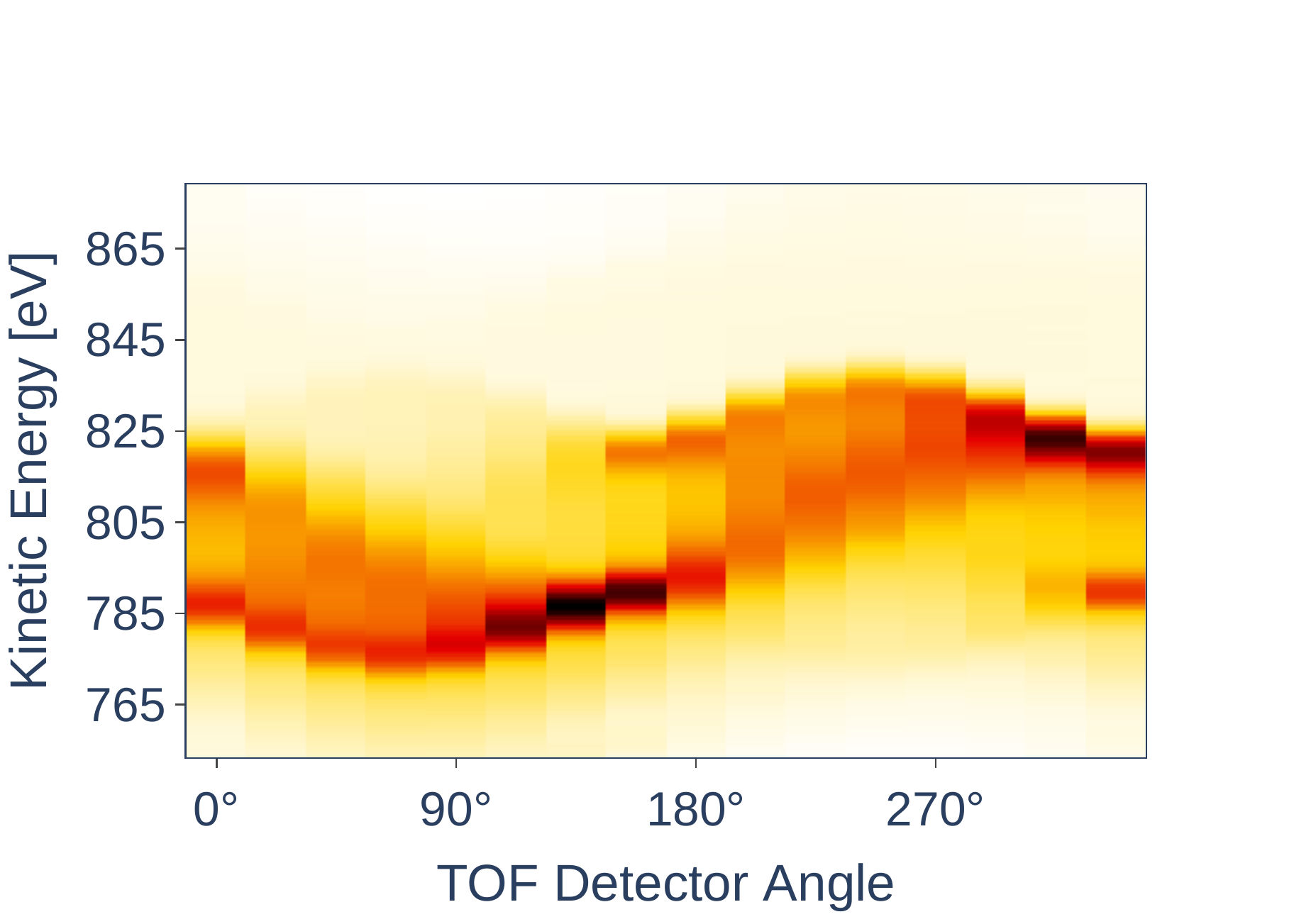}\label{fig:decayDetectorImage}}
    \caption{Experimental (a) and simulated (b)--(h) data for an angular streaking measurement. (a) shows real data from Hartmann \textit{et al.}~\cite{hartmann2018attosecond} in neon. (b) displays a simulated detector image under similar conditions with the addition of $\pm30~\%$ noise. In [(c) and (f)] a simulated spectrogram and the corresponding detector image, respectively, are shown before  and after [(d) and (g)] adding a pulse structure to a Gaussian intensity distribution. Note the very slight differences between (f) and (g), which are caused by the additional SASE pulse structure and give rise to an accordingly spiky pulse reconstruction. (e) and (h) display a simulated Auger electron spectrogram and the respective angularly streaked detector image, after ionization of the Ne 1s shell with an arbitrarily chosen decay time of 7.6 fs. All displayed figures are normalized to the interval $[0,1]$.}
    \label{fig:spectrogram-detectorImage}
    \end{figure*}

\section*{Application Case: Angular Streaking}

    A long-standing goal in laser and X-ray research is to enable measurements providing both temporal and spatial real-time information about electronic and consequential structural changes on a molecular level with element-site specificity---a so-called \textit{molecular movie}. For this, a suitable ultrashort X-ray pulse duration is one of the key parameters, which is both hard to facilitate and difficult to measure. Yet, a reliable time-resolving experimental method is essential for determining parameters such as the detailed intensity profile of the SASE FEL pulse \cite{Bonifacio1994}, corresponding damaging thresholds for materials under investigation \cite{Neutze2000}, the nanoscale interpretation of ultrafast single-shot diffraction imaging \cite{Barty2008}, and the probabilities for multi-photon processes \cite{Young2010, Rudenko2017}, to name a few.
    Applying the \textit{angular-streaking technique} \cite{eckle2008attosecond} to the field of XFELs leverages a versatile approach for the temporal and spectral characterization of individual (X)FEL pulses \cite{hartmann2018attosecond}.
	
	The applied scientific instruments for this new method are \textit{angle-resolving electron spectrometers} \cite{hartmann2018attosecond, duris2020tunable}. In case of the first demonstration of angular streaking at XFELs and also in the present case, 16 individually working time-of-flight (TOF) spectrometers are arranged in a ring-like structure around the target region, perpendicular to the propagation direction of the incoming X-rays \cite{hartmann2018attosecond}. Together with a co-propagating circularly polarized infrared laser, spatially and temporally overlapped with the XFEL at the target region, this setup enables angular streaking. Atoms from a target gas are ionized with the XFEL pulse and the emitted electrons are swept, i.e., streaked, in energy \textit{and} angle by the concomitant rotating electric field vector of the infrared laser. In a simplified picture, the streaking field vector can be understood as the hand of a clock that encodes the parameter time via the angles at which electrons are detected with accordingly shifted energies (see illustration in Fig.~\ref{fig:ga}). Given sufficiently many electrons to "report" on their ionization time within the SASE pulse, the measured electron emission patterns contain the information of the full time--energy structure of the ionizing XFEL pulse with attosecond resolution. The method can be adapted for pulses with different photon energy by selecting target gases of suitable electron binding energies and photoionization cross sections. The mechanism and experimental setup for the angular-streaking technique as applied to SASE X-ray pulses is described by Hartmann et al.\cite{hartmann2018attosecond} and the general principle can be found here \cite{Constant1997,Itatani2002,Kazansky2016}.
	
	 In the experiment under consideration, a time trace is measured for photoelectrons emitted by each X-ray shot and in each TOF spectrometer, hence, generating 16 traces at a rate set by the repetition frequency of the XFEL and of the overlapped streaking laser. For single-shot spectroscopy, a trace represents the number of electrons arriving after specific flight times.
	 These time-domain traces can be converted to the energy domain (\textit{spectra}) by taking into account the length of the flight path and the actions of additional electric fields along their paths, which are routinely used for enhancing the achievable energy resolution and the electron collection efficiency. The combined representation of a full angle-resolved streaking measurement forms an image with $16$ columns, representing the respective detector angles, and several rows corresponding to the range of electron energies detected in the specific measurement (\textit{detector image}, cf.\ Fig.~\ref{fig:real_shot}). Time-dependent electron spectra (\textit{spectrograms}) are then generated by converting the emission angles to times using the known rotation period of the electric field vector for the circularly polarized streaking laser. (cf.\ Figs.~\ref{fig:perfSpectrogram} \& \ref{fig:pulseSpectrogram}). In the present ML case study, we have simulated the spectrograms and the according detector images based on the equations for streaking previously derived \cite{Itatani2002}, following the procedure established and described in detail in the SI of Hartmann et al.\cite{hartmann2018attosecond}. Thus, we can at will generate both a huge "data" set for training the ML algorithms, as well as an additional set of completely known target shots for testing the developed NN predictions.
	 
	 For an X-ray pulse with no temporal and spatial overlap in the interaction region with an external streaking field (\textit{unstreaked shot}), the spectra in all detectors are showing the characteristic electron energy distributions (\textit{spectral lines}) for the target under investigation.
	 Typically each line also shows an angular dependence in signal intensity. 
	 In Figs.~\ref{fig:perfDetectorImage} \& \ref{fig:pulseDetectorImage} one can see this variation in the low intensity regions around $0^\circ$ and $180^\circ$ in contrast to the high-intensity parts at $90^\circ$ and $270^\circ$, with intermediate intensities in the columns at angles in between.
 	 
 	  If a \textit{circularly polarized streaking laser} is present, the detector image is modulated according to the instantaneous streaking laser vector potential, leading to a sinusoidal variation of the spectral lines along the angle axis.
 	  
 	 \begin{figure}[!htbp]
	    \centering
	    \includegraphics[width=0.5\textwidth]{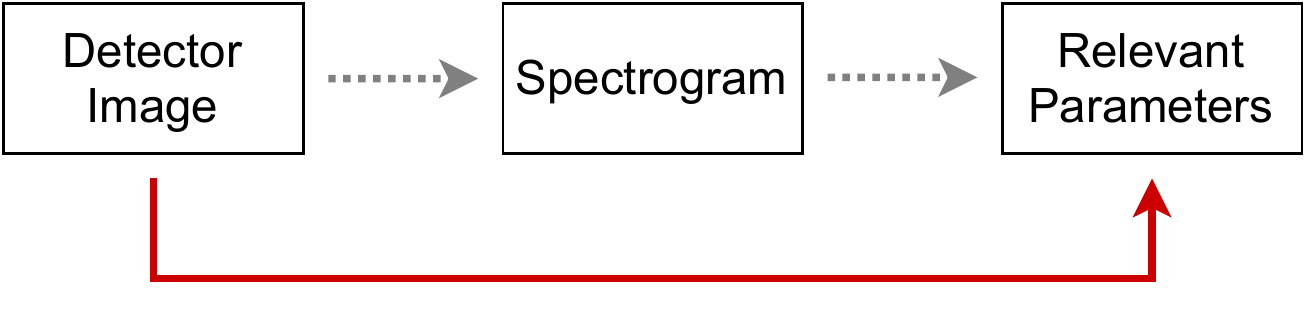}
	    \caption{There are two ways to extract important pulse characteristics from the detector image. The dotted line refers to a full reconstruction (spectrogram) and the extraction of relevant parameters out of this reconstruction. The other way is to skip the full reconstruction and to concentrate on typically most relevant parameters only. In this article, we choose the second approach.}
	    \label{fig:pipeline}
	\end{figure}
 	  
   The goal regarding SASE FEL X-ray pulse characterization and their potential control is to reconstruct the spectrogram from a measured detector image, which gives the full information about the X-ray time--energy structure (cf.\ Fig.~\ref{fig:pipeline}, dotted line).
	 In many experimental situations, however, it is sufficient to restrict our analysis to some of the \textit{most relevant} SASE X-ray parameters (cf.\ Fig.~\ref{fig:pipeline}, red line).
	 In the subsequent discussion, we have, therefore, focused on NN predictions of the temporal aspects of ultrashort FEL pulses. Details about the NNs' framework conditions, the chosen architecture, and hyperparameter optimization can be found in the Methods below.
	
	We picked the following pulse characteristics for a comparison of their reconstruction by the NNs (\textit{prediction}) with the originally simulated data (\textit{target}):
	
    \paragraph{\textbf{Kick}}
    The kick is the maximum streaking shift in electron kinetic energies for each X-ray shot, and thus for a given temporal delay and phase relation between the X-ray pulse and the infrared streaking laser. There are two main reasons for a change in the kick from shot to shot. The first is the \textit{relative timing jitter} between the X-ray and the streaking pulse \cite{Bionta2011,Hartmann2014,Diez2021}, which is unavoidable due to the stochastic generation process of the SASE mechanism and additional fluctuations in arrival time caused by air fluctuations, thermal expansion in optomechanical components, and general synchronization errors between the two separate laser pulses. 
    The second reason for variations of the kick is the \textit{random change of the carrier--envelope phase} of the streaking laser from shot to shot. One can solve this by stabilizing the carrier--envelope phase \cite{Fuji2005}, which is a rather difficult technical requirement, or by using the technique of angular streaking \cite{hartmann2018attosecond}, which is the basis for the simulations studied in this article.
    
    Since we translate a temporal distribution (X-ray intensity structure) into an energy distribution (kinetic energy of the streaked photoelectrons), a lower kick value means a more shallow gradient of the streaking ramp, which is given by the kick over the cycle period of the electric field corresponding to the streaking laser wavelength. Thus, the resolution of the measurement is directly degrading with decreasing kick. In this sense, the determination of the kick strength is not so much interesting in itself, but is a measure for the quality of the reconstruction and can be used as a filtering handle of the data. It is also a good consistency check for the functioning of the applied NNs, since the kick is a parameter that can also be readily assessed with other analysis methods.
    
    \begin{figure}[!htbp]
	    \centering
	    \includegraphics[width=0.6\textwidth]{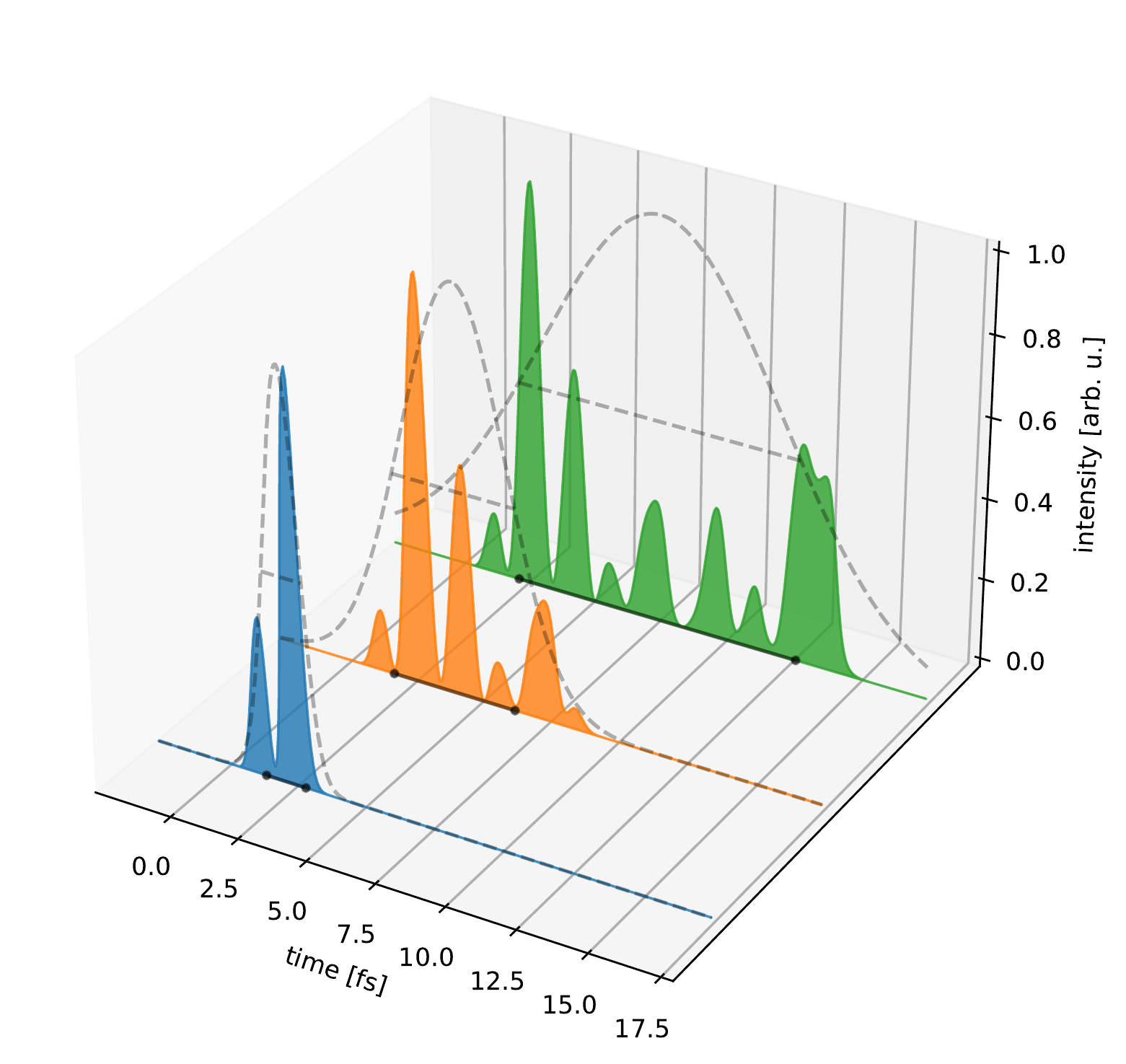}
	    \caption{Three examples of different SASE XFEL pulse intensity structures showing varying total duration and complexity generated with OCELOT~\cite{ocelot}. The pulse durations given as FWHM/RMS, respectively, for the exemplary shots are 1.5 fs/640 as (blue), 4.5 fs/1.9 fs (orange) und 10.4~fs/4.4 fs (green). The corresponding Gaussian pulse envelopes are shown as dashed grey curves, including the FWHM durations, and their projections onto the time axes as black lines.}
	    \label{fig:pulseStructure}
	\end{figure}
    
    \paragraph{\textbf{Pulse Duration}}
    The pulse duration is the most important parameter for many ultrafast free-electron laser experiments, e.g., a variety of pump/probe measurements of electronic state changes or investigations of nonlinear excitation dynamics~\cite{Rudenko2017, rudek2012ultra, Picon2016, li2021electron}, albeit it is one of the most difficult to measure directly.
    Especially for XFEL SASE pulses, each pulse has a different duration and erratic intensity structure that even complicates the definition of the term \textit{pulse duration}.
    In this article, we use the \textit{root-mean-square (RMS) duration}, i.e., the square root of the time variance of the temporal intensity profile~\cite{sorokin2000measurement},
    \begin{equation}
        t_{\rm p,RMS} =
        \sqrt{\langle t^2 \rangle - \langle t \rangle^2},
        \label{eq:rmsduration}
    \end{equation}
    where
    \begin{equation}
        \langle t^n \rangle =
        \frac{1}{N} \int_{-\infty}^{\infty} t^n I(t) \; dt \quad \text{and}
        \quad
        N = \int_{-\infty}^{\infty} I(t) \; dt
    \end{equation}
    are the $n$-th moment and the normalization constant, respectively, as the basic definition of the \textit{pulse duration}.
    
    A common choice for more well-behaved Gaussian-like laser pulses from table-top systems is the \textit{full width at half-maximum (FWHM)}. Given a Gaussian distribution with standard deviation $\sigma$, corresponding to the RMS duration in this case, the FWHM is calculated as follows: 
    
    \begin{equation}
    \label{eq:fwhm_duration}
        FWHM = 2 \sqrt{2\ln2}\sigma \approx 2.35 \cdot \sigma.
    \end{equation} 
    
    As SASE pulses are generally spiky and irregular (cf. Fig.~\ref{fig:pulseStructure}), this metric is not fully applicable. In our simulation case, however, this quantity is nevertheless of interest due to the fact that we use the FWHM to generate (and in fact define) the Gaussian distribution envelopes in Fig.~\ref{fig:pulseStructure} and because the FWHM better relates to the intuitive concept of a full-length pulse duration.
    The RMS duration, however, gives a more complete measure of the temporal distribution of the pulse energy including possible pulse wings~\cite{sorokin2000measurement} or substructure (see also the next paragraph).
    
    \paragraph{\textbf{Pulse Structure}}
    Due to the microbunching in the FEL each SASE pulse has an individual intensity profile, made up of several shorter ‘spikes’ with random intensities (cf. Fig.~\ref{fig:pulseStructure}). The average number of spikes per pulse is determined by the specific operation parameters of the XFEL. It can be expressed in a statistical treatment as the number of individual energy modes contributing to the XFEL pulse~\cite{Krinsky2003}. The ensuing pulse shape can be arbitrarily complex. The shorter the overall pulse duration in relation to the single-spike length, i.e., the fewer spikes per complete pulse, the more important individual spikes are becoming (Fig.~\ref{fig:pulseStructure}). Especially for estimating the damage thresholds of investigated probes as well as for experiments sensitive to the instantaneous X-ray intensity, or for ultrafast pump/probe measurements, the XFEL pulse structure needs to be known exactly to interpret the observed data on a shot-to-shot basis unambiguously.
    
    \paragraph{\textbf{Auger Decay Time}}
    Many of the scientifically interesting processes of non-equilibrium physics and structure-changing chemistry are not directly triggered by the exciting X-ray pulse but are the result of subsequent complex relaxation dynamics. These dynamics are determined by the time-dependent, i.e., transient electronic structure of the system under study. One of the most fundamental electronic processes after inner-shell ionization of matter by X-rays is the \textit{Auger decay}, whereby a second electron from an outer shell fills the generated core hole and transfers the excess energy to a third electron (\textit{Auger electron}), which is then emitted from the ion. 
    This process is specific to the contributing discrete electronic states of an atomic or molecular system and has a characteristic time constant for the emission of the third electron (\textit{Auger decay time}). In our simulations, we assume that one Auger decay channel dominates for neon (Ne) after 1s ionization. The corresponding Auger decay time on the order of 2~fs to 3~fs \cite{Haynes2021} can serve as a fundamental benchmark for demonstrating the capability of the method to retrieve ultrafast timing information from recorded data.

\section*{Results}\label{sec:Results}

\begin{figure*}
    \centering
    \vspace{-3em}
    \begin{minipage}{.49\textwidth}
    \subfloat[Difference between the target and the predicted kick label, demonstrated on a test set with $4 \cdot 10^{4}$ samples. 96~\% of all predictions have a smaller deviation than 10~\% of the respective target value.]{\includegraphics[width=\textwidth]{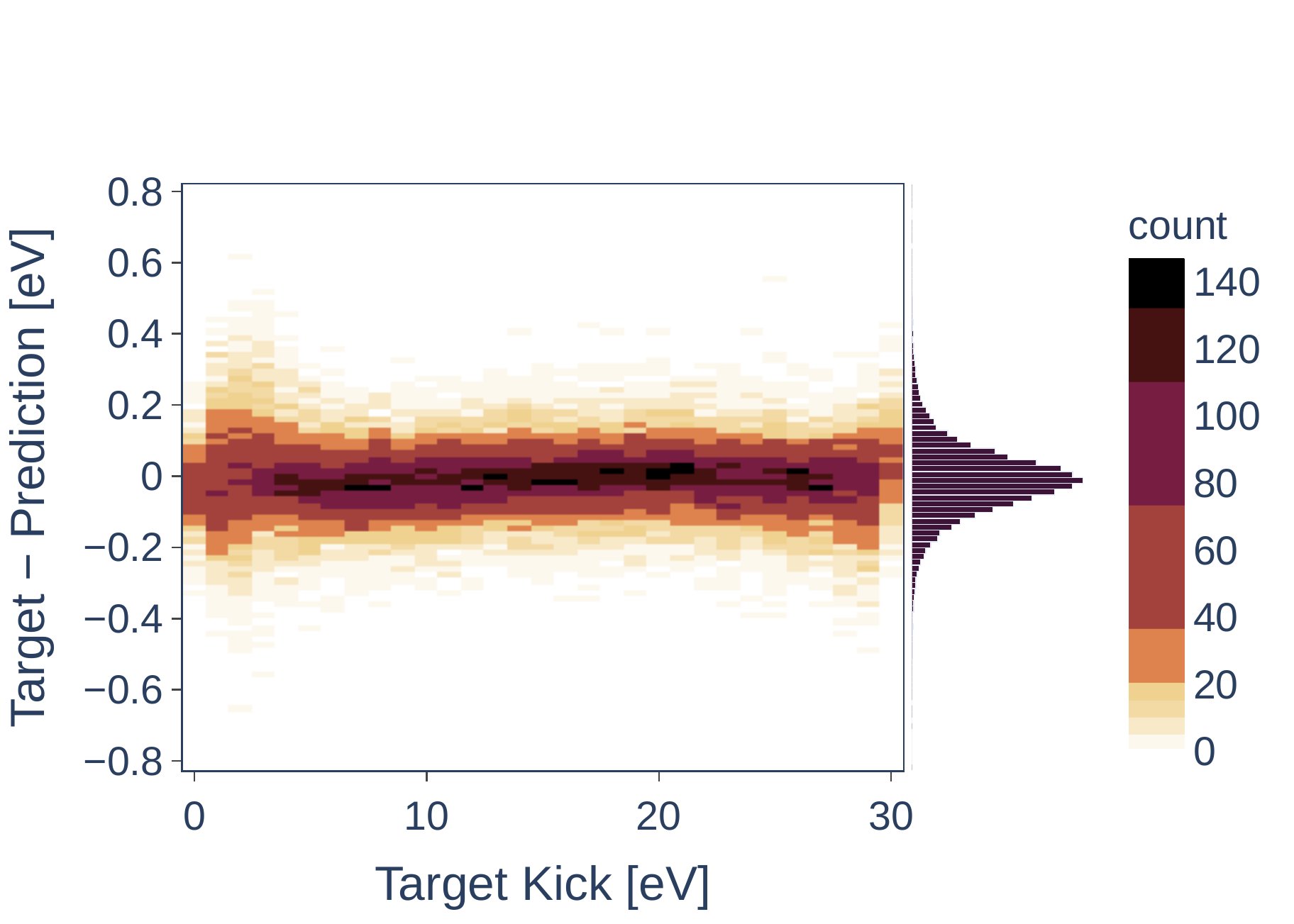}\label{fig:kick_res}}
    \vfill
    \subfloat[Difference between the target and the predicted FWHM pulse duration label, demonstrated on a test set with $4 \cdot 10^{4}$ samples. In 94~\% of all cases where the kick was greater than $5$~eV, the deviation between prediction and target value was less than 1~fs.]{\includegraphics[width=\textwidth]{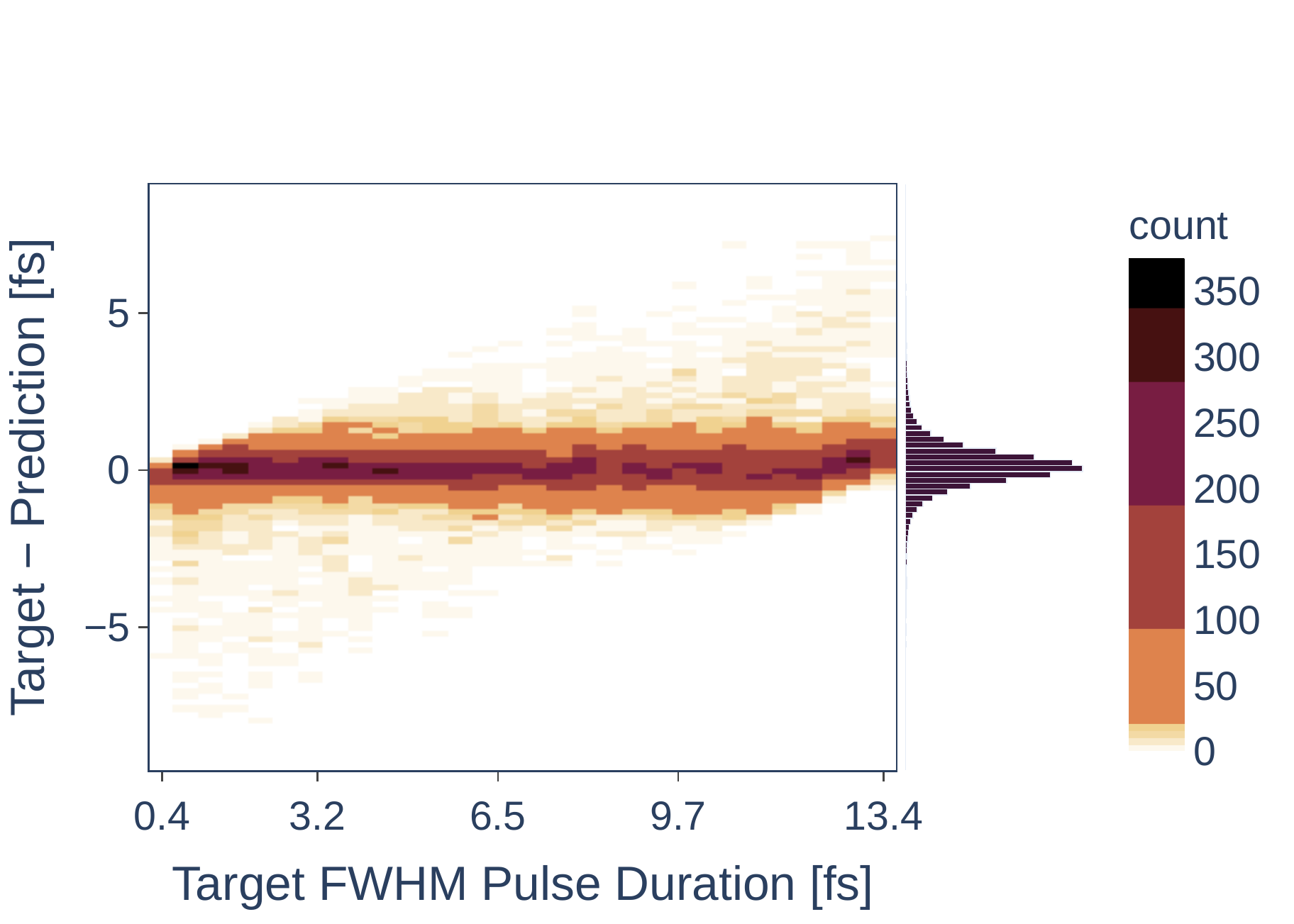}\label{fig:pl_res}}
    \vfill
    \subfloat[Difference between the target and the predicted Auger decay time, demonstrated on a test set with $4 \cdot 10^{4}$ samples. In 92~\% of all cases where the kick was greater than 3~eV, the deviation between prediction and target value was less than 0.5~fs. ]{\includegraphics[width=\textwidth]{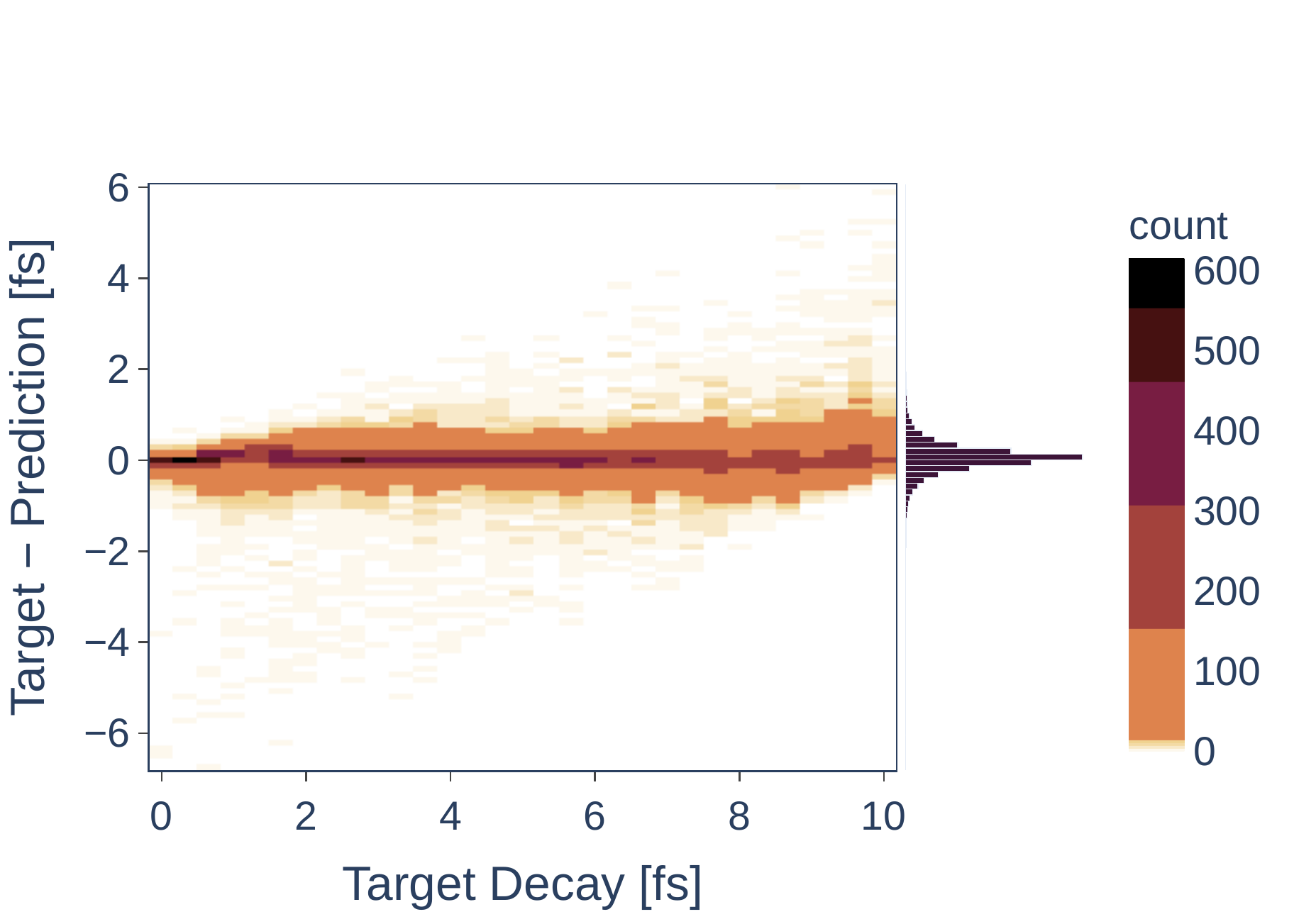}\label{fig:decay_res}}
    \end{minipage}\hfill
    \begin{minipage}{.49\textwidth}
    \subfloat[Deviation of the computed RMS pulse duration from the predicted pulse structure to the RMS pulse duration of the target pulse structure.]{\includegraphics[width=\textwidth]{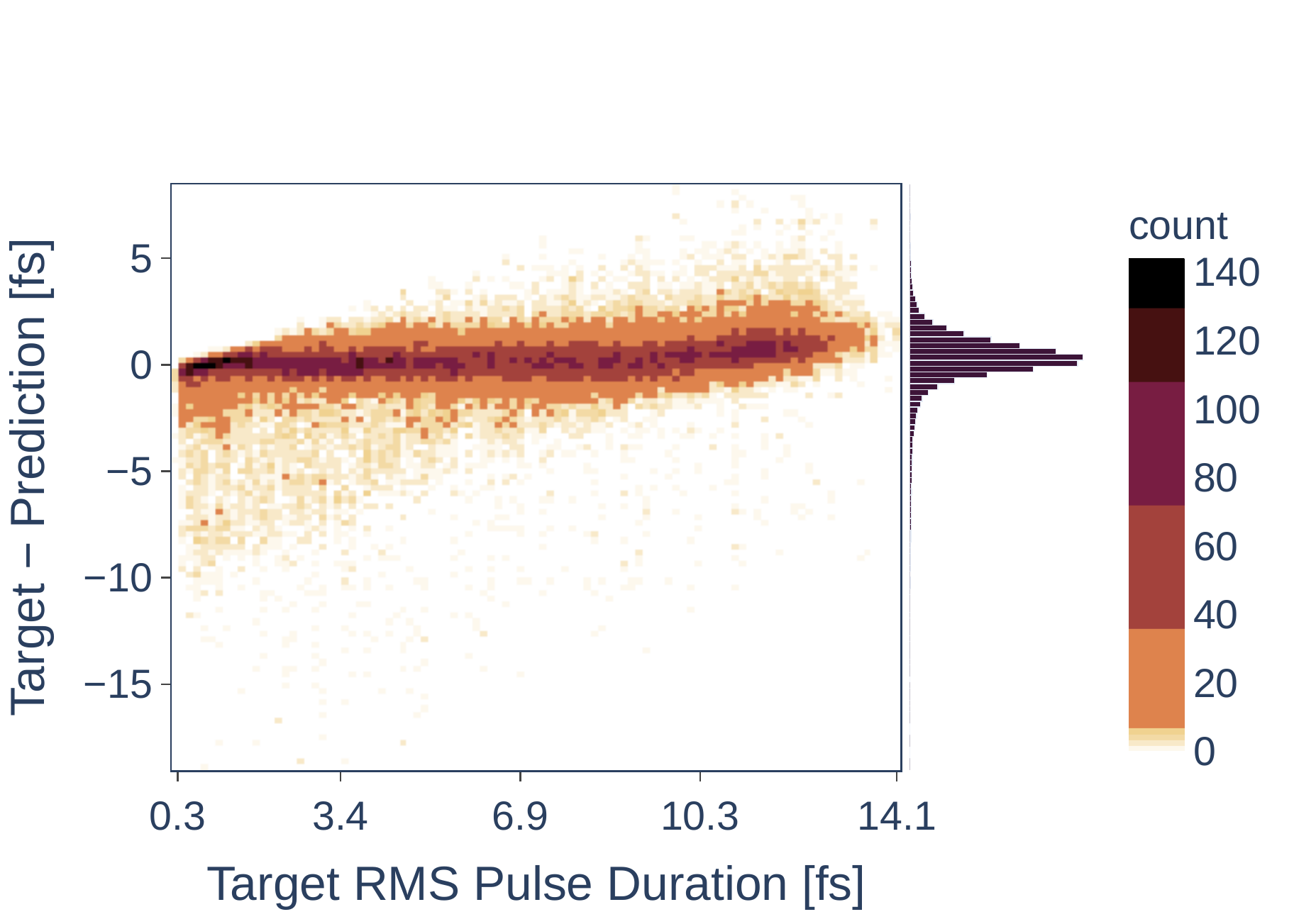}\label{fig:rms_pl_comp}}
    \vfill
    \subfloat[Comparison of the FWHM pulse duration prediction as a function of energy kick. The accuracy of the pulse length estimate depends on the respective kick value of the shot. The higher the kick, the more accurate the prediction of the FWHM pulse duration gets. ]{\includegraphics[width=\textwidth]{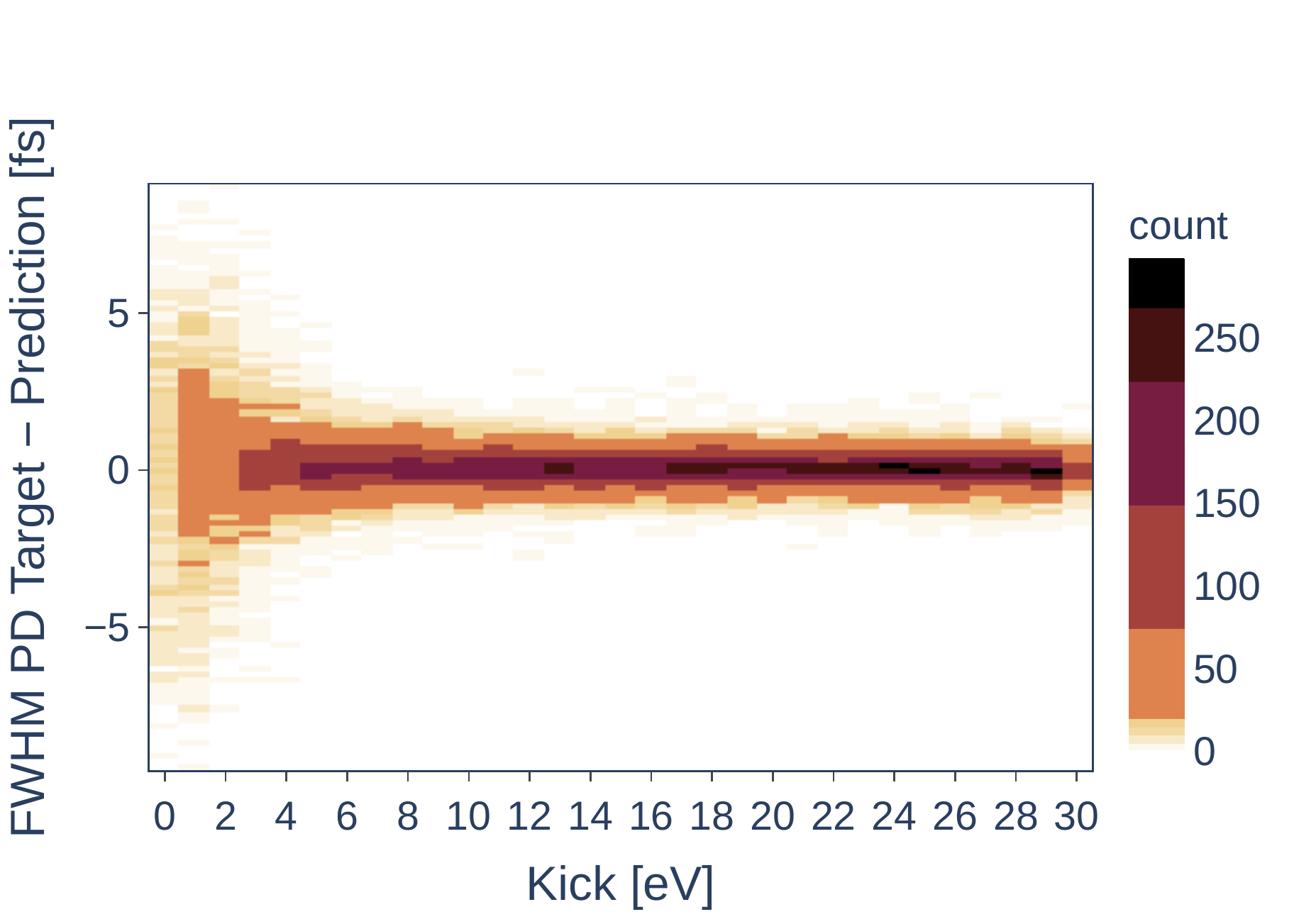}\label{fig:pl_kick}}
    \vfill
    \vspace{-1em}
    \vfill
    \subfloat[Comparison of the decay prediction as a function of energy kick. The accuracy of the Auger decay time estimate depends on the respective kick value of the shot. The higher the kick, the more accurate the prediction of the Auger decay time gets.]{\includegraphics[width=\textwidth]{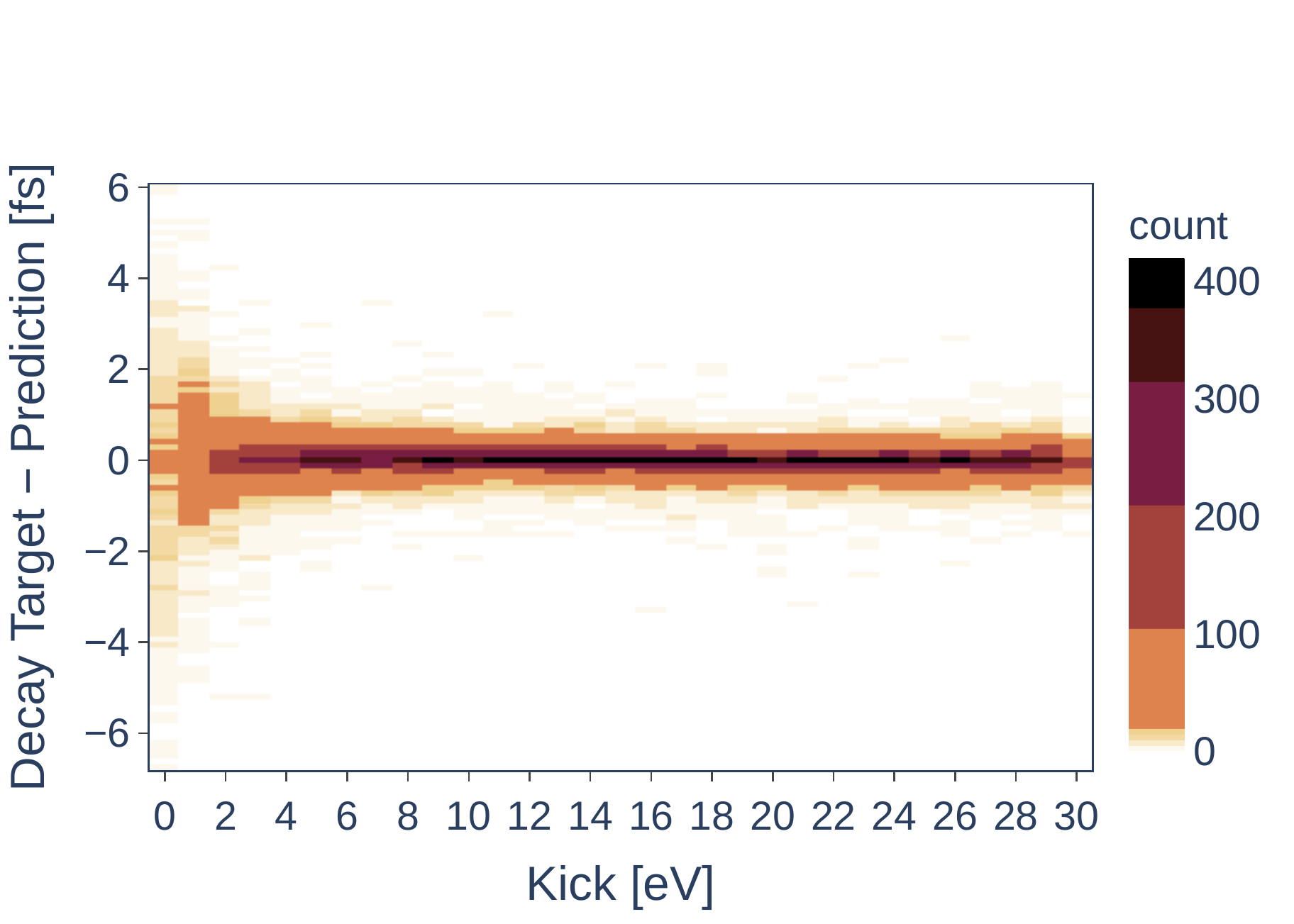}\label{fig:decay_kick}}
    \end{minipage}
    \caption{Prediction accuracies and dependencies of the labels \textit{kick}, \textit{RMS pulse duration}, \textit{FWHM pulse duration}, and \textit{decay} using data sets containing all levels of noise [$\pm 0~\%$, $\pm 10~\%$, $\pm 20~\%$, $\pm 30~\%$]. The slight skew of the distributions in Figs. (b), (c) \& (d) stems from the inherent tendency of NNs to predict values closer to the mean of the learned parameter space for unclear inputs, i.e. from detector images with too low kicks.}
    \label{fig:results}
    \end{figure*}
	
	All of the above described SASE XFEL pulse characteristics can be predicted with varying degrees of accuracy by utilizing convolutional NNs.
	For each pulse characteristic, we will examine the results of the trained models in more detail.
	
	\subsection*{Kick}
	
	Of all the characteristics studied, the kick turned out to be the easiest to predict. Fig.~\ref{fig:kick_res} shows that most of the predictions only slightly deviate from the respective targets. It turns out that $96~\%$ of all predictions have a smaller deviation than $10~\%$ of the respective target value.
	Though the kick can easily be derived from the detector images, 
	an accurate estimate of this parameter is necessary for better judging the reliability of the reconstruction for the \textit{FWHM pulse duration}, \textit{pulse structure} and \textit{Auger decay time}. This will become apparent in the following paragraphs.
	
	\subsection*{\label{subsec:PD}FWHM Pulse Duration}
	
	The comparison of predicted and target FWHM pulse durations is shown in Fig.~\ref{fig:pl_res}. As for the kick estimates, the majority of the values are well reconstructed. However, it is evident that a few of the predictions deviate strongly from the target values. One hypothesis explaining this behavior is that for smaller kicks the resolution of the measurement degrades and predicting a pulse duration can become arbitrarily difficult. That is why we have investigated the accuracy of the FWHM pulse duration estimate against the true kick value.
    
    Fig.~\ref{fig:pl_kick} confirms the previously stated hypothesis. Above a (true) kick value of approximately $5$ eV, estimating the FWHM pulse duration becomes feasible. In fact, in $94~\%$ of all cases where the kick was greater than $5$~eV, the deviation between prediction and target value was less than $1$~fs. This is due to the fact that small kick values on the order of the nominal SASE bandwidth correlate to unsuccessful angular streaking shots that need to be discarded anyway. Supplementary Figs.~2a~and~2b
    display exemplary shots with a large and a small kick, respectively.
	
	\subsection*{Auger Decay Time}
	The auger decay time can also be well approximated by the respective NN (cf. Fig.~\ref{fig:decay_res}). Most of the estimates hardly deviate from the zero line of the difference between the target value and the prediction. However, as for the FWHM pulse duration estimates, there are some outliers strongly deviating from the true Auger decay time value. Using the same reasoning as for the FWHM pulse duration estimates, we have compared the prediction of the Auger decay time value with the true kick value.
	Here, the same behavior can be observed as for the FWHM pulse duration (cf. Fig.~\ref{fig:decay_kick}). It is evident that a reasonable determination of the Auger decay time is only possible for a kick value of $3$~eV or higher. In fact, in $92~\%$ of all cases where the kick was greater than $3$~eV, the deviation between prediction and target value was less than $0.5$~fs.
    It follows that shots with small kick values should be discarded in advance to appropriately approximate the true Auger decay time. As before, Supplementary Figs.~2c~and~2d
    display exemplary shots with a small and a large kick for the decay reconstruction, respectively.
    
    \subsection*{Pulse Structure and RMS Pulse Duration}\label{subsec:pulse_structure}
	
\begin{figure*}[!htbp]
    \centering
    \subfloat[Simple SASE pulse structure.]{\includegraphics[width=0.49\textwidth]{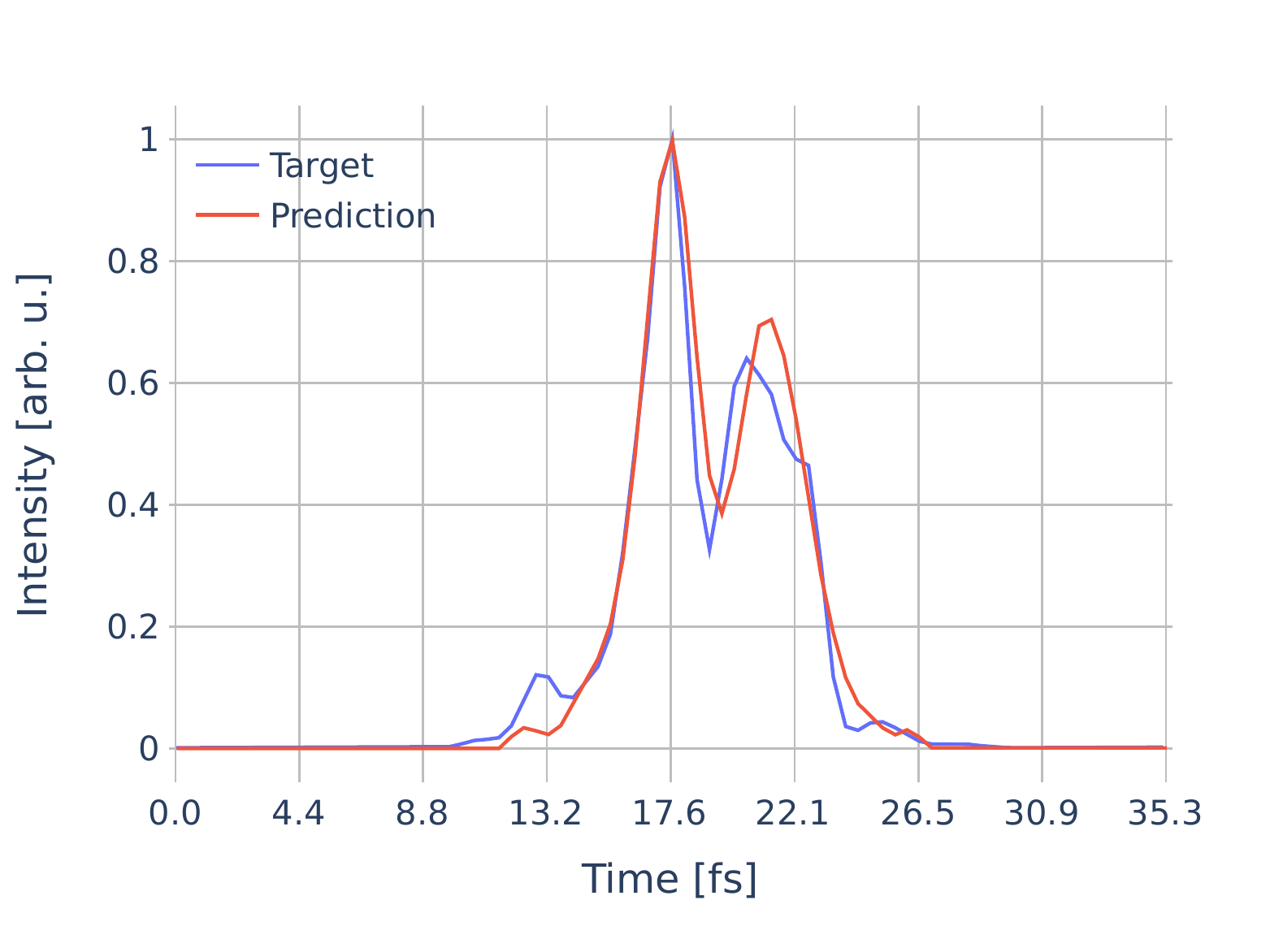}\label{fig:8a}}
    \hfill
    \subfloat[Complex SASE pulse structure.]{\includegraphics[width=0.49\textwidth]{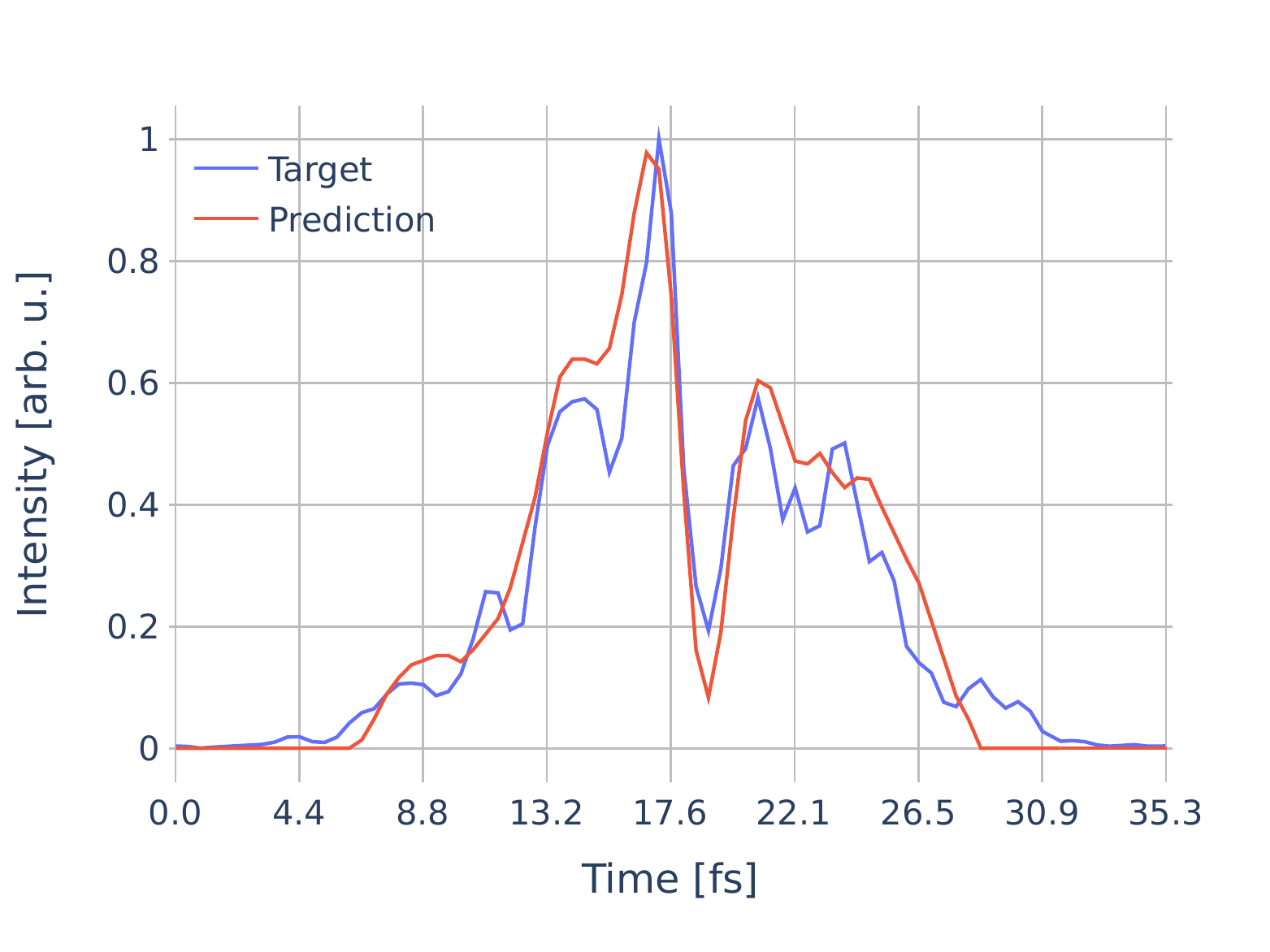}\label{fig:8b}}
    \caption{Examples for simple (a) and more complex (b) simulated and reconstructed SASE pulse structures with RMS pulse durations of $6.2$~fs and $11.7$~fs, respectively.}
    \label{fig:TDexamples}
    \end{figure*}

	The full temporal pulse structure of the SASE pulse is probably the most difficult property to predict in our study.
	This is not surprising, as it is also the most complex one, being represented by a vector instead of a single value in case of the kick or the pulse duration, which holds the information of the intensity distribution over time.
	Altogether, the trained network works relatively well in its objective to predict the trend, i.e., peak positions and their relative intensities of the pulse structure.
	However, as has been expected, these predictions get less accurate for more complex pulse structures. This behavior can be seen for two different exemplary simulated SASE pulses in Fig.~\ref{fig:TDexamples}, one relatively simple \ref{fig:8a} and one more complex \ref{fig:8b}, in which not all of the finer structures could be reliably reproduced. Nevertheless, the main features including the larger peaks can always be predicted.
	
	Remarkably, the quality of the predicted pulse structure does not show a significant dependency on the value of the kick; except for a kick very close to or equalling zero, which is not surprising, as this again corresponds to an unsuccessful event where basically no streaking occurred.
	There is also no significant dependency on the duration of the pulse, as one might have expected. An absolute value for the mean squared error (MSE) does indeed increase with the pulse duration; normalized to it, however, the average `MSE per time step' is more or less constant (cf. Supplementary Fig.~3). 
	Now that our model is capable of extracting the pulse structure, it is possible to compute the RMS pulse duration by using Eq.~\ref{eq:rmsduration}. Fig.~\ref{fig:rms_pl_comp} shows the deviation of this computed RMS pulse duration to the RMS pulse duration of the target pulse structure.
    The average deviation lies below $1$~fs. Only for very long pulse durations there is a trend towards a slight underestimation, although the error remains just around 10~\% in most cases.
    We note that an additional NN for the direct prediction of the RMS pulse duration could serve as a comparative measure for the RMS value calculated from the pulse structure, allowing a coarse estimation of the quality of the reconstructions.
    
    \subsection*{Influence of Noise on the NN Performance}
    
    In order to investigate the influence of noise on the predictions, we generated a test set comprising of detector images with several different noise levels ($p = [0.0, 0.1, 0.2, 0.3]$) as shown in Eq.~\ref{eq:noise} in the Methods section and fed it into the NNs.
    
    As expected, additional noise influences the result of the NN's prediction (cf. Tab.~\ref{tab:variance}).
    An example for a noisy simulated detector image is given in Fig.~\ref{fig:noiseLevels}, more expressive examples for different simulation settings are shown in Supplementary Fig.~4. 
    The prediction for non-noisy data is nearly perfect, whereas the prediction for noisy data slightly differs from the target.
    Despite the decreased prediction accuracies, it is evident that the NNs can handle noise robustly.
   
    \begin{table}[!htbp]
    \centering
    \begin{tabular}{l|lllll} 
    \toprule
    Noise & Kick & FWHM & RMS & Decay \\
    \midrule
    $\pm  0 \%$ & $0.19$~eV & $0.52$~fs & $0.67$~fs & $0.08$~fs \\ 
    $\pm 10 \%$ & $0.24$~eV & $0.88$~fs & $0.65$~fs & $0.17$~fs \\
    $\pm 20 \%$ & $0.34$~eV & $1.05$~fs & $0.65$~fs & $0.26$~fs \\
    $\pm 30 \%$ & $0.43$~eV & $1.2$~fs  & $0.73$~fs & $0.34$~fs \\
    \bottomrule
    \end{tabular}
    \caption{Standard deviations of the predicted labels with regard to the respective targets kick, pulse duration, and Auger decay computed on $1000$ samples, respectively. 
    Settings for the kick, FWHM/RMS pulse duration, Auger decay sample data:
    \textit{kick}~=~$22.5$~eV, \textit{pulse duration}~=~$4.85$~fs, \textit{Auger decay} = $7.0$~fs.}
    \label{tab:variance}
    \end{table}

    \section*{Discussion \& Outlook: Online SASE-Pulse Characterization and Shaping} \label{sec:FutureWork}

    So far, we have shown that several characteristics of XFEL pulses are predictable with varying degrees of accuracy. To investigate how close the current status comes to real-time pulse characterization during experimental campaigns, we need to address a number of different issues, which can each be tackled exploiting the specific analytical strength made accessible by the methods of NNs:
	
	\paragraph{\textbf{Output Speed}}
	
	For evaluating the input images at full speed of the XFEL repetition rate in the kHz to MHz regime, an efficient analysis is inevitable. NNs are known for delivering outputs quickly. We have used several batch sizes, starting from one image up to $4096$ as input. Investigating this is important as such a comparison determines whether a batch-wise and therefore highly parallelized analysis is performing better than analyzing image-wise. Batch-wise evaluation is specifically suited for the European XFEL facility, since a train with a very fast succession of pulses (maximum 2700 per 600 $\mathrm{\mu}$s) is followed by a pause of several milliseconds that can be used for analysis purposes \cite{decking2020mhz, lederer}. We have tested how fast the NN output is generated on a GeForce RTX 2070 GPU using single precision floating point format (Tab.~\ref{tab:speed}).
	
	\setlength{\tabcolsep}{3.8pt}
    \renewcommand{\arraystretch}{1.5}
	
	\begin{table}[!htbp]
	\centering
    \begin{tabular}{lllllllll}
    \toprule
    \textbf{BS}        & $1$    & $64$   & $128$  & $256$  & $512$  & $1024$ & $2048$ & $4096$ \\
    \midrule
    \textbf{Dur [ms]} & $1.38$ & $1.44$ & $1.48$ & $1.54$ & $1.62$ & $1.54$ & $1.38$ & $1.44$ \\
    \bottomrule
    \end{tabular}
    \caption{Time measurements for predictions of the trained model on a GeForce RTX 2070 card with different batch sizes (BS). We ran hundred experiments and averaged the results.}
    \label{tab:speed}
    \end{table}
    
    The model is able to reach quick predictions mostly independent of the batch size as the computation on a GPU runs all tasks, i.e., computes a prediction for each image within the batch, in parallel. In general, the number of input images in one batch is only limited by the RAM of the used GPU. Thus, it is apparent that it is advantageous to analyze a larger batch of data than individual images. With a batch size of $4096$ our current model is already able to keep up with the European XFEL in high-repetition mode for online predictions.
    
    \paragraph{\textbf{Reliability Estimation}} 
    
    Next to fast evaluation, a degree of certainty in the NN predictions must be ensured. As shown in the results, the NN prediction may deviate quite substantially from the target. Some of the difficulties can be directly circumvented. By determining the kick, for example, we can already filter whether a prediction regarding the labels \textit{Auger decay time} or \textit{pulse duration} is reasonable. But this still does not give us a direct statement specifying how certain the NN's prediction is. Optimally, we would want to have a reliable measure of how good the predictions of the trained models are, even for unknown shots without a target. 
	
	There are several ways to determine the prediction uncertainties of NNs. The epistemic uncertainty determines the uncertainty due to insufficient knowledge. This can be implemented, for example, via Monte Carlo dropout \cite{Gal2016} or Monte Carlo batch normalization \cite{Teye2018}. The aleatoric uncertainty determines the uncertainty due to the complexity of the problem, and can be investigated by creating a fitted cost function \cite{kendall2017uncertainties}. This topic is currently a very active area of ML research. We are in the process of developing our own approach to benchmark the specific task of stochastic X-ray pulse reconstruction, ideally combining both uncertainty determinations in one procedure as previously shown\cite{kendall2017uncertainties}.
	
	\paragraph{\textbf{Gap between Simulation and Reality}}
	
	So far, our NNs are suited only for data that look exactly like the input shown in Fig.~\ref{fig:spectrogram-detectorImage}.
	Whether the NNs are suitable for predictions on experimental data (cf. Fig.~\ref{fig:real_shot}) is not easy to validate, especially since we do not have true labels for the experimental data. In addition, we need to identify to what extent our modeled noise replicates real noise of the spectrometer, e.g., electronic ringing of the detector readout or background signals from undesired processes. There are two ways to tackle the gap between simulation and experiment.
	Either the real data must be denoised before the analysis (e.g., Denoising Autoencoders~\cite{Vincent2010}) or the simulation data must be provided with additional, appropriately modeled noise. Simultaneous approaches in both directions should give a more complete understanding for mitigating this issue in future efforts.
	
	\paragraph{\textbf{Responding to Changes}} 
	
	We have shown that our developed NNs work on data with several levels of noise.
	However, when utilizing real-life TOF spectrometers, it may occur that TOF sensors fail or produce unrealistic results. In such cases NN re-training or knowledge extension is inevitable. Here, online learning \cite{Fontenla-Romero2013} is a helpful tool. In this case, the model is trained continuously on newly generated data. Thus, the training can be quickly adapted to new environments. To circumvent the catastrophic forgetting \cite{Kirkpatrick3521} of NNs, continual learning \cite{PARISI201954,He2021} may be utilized.
	
	\paragraph{\textbf{Pulse Shaping}}
	
	The term pulse shaping can be interpreted in two technically different ways. With the online analysis methods already demonstrated in the current manuscript, we can establish an X-ray sorting scheme based on the full characterization of every single pulse and the possibility to filter for desired pulse shapes and durations. Especially with high-repetition rate XFELs, this can be a vital form of “passive shaping”. The second, and more exciting, route refers to an actual pulse shaping in terms of intelligent experimentation schemes and dynamical interaction of the machine with simultaneous photon-based measurements. Thus, the live updates on X-ray pulse changes may be used for a more detailed control of the parameters and for actual SASE pulse shaping. First steps for this interaction have been identified to entail a feedback loop to the accelerator that provides an online data stream of, e.g., X-ray pulse duration and spectrum. The machine operators can choose to engage this loop into their electron bunch compression algorithms with a preset goal of optimization to be pursued. Further steps towards intelligent and active experimentation are in development and will be presented in future studies.
	
	\paragraph{\textbf{Conclusion}}
	
	In this article, we demonstrated a path toward online characterization of free-electron laser pulses by applying NNs on detector images captured with angular streaking. In addition to several predictable characteristics, we have been able to identify and confirm dependencies between the respective characteristics that can be used to control the machine settings during experimental campaigns. This way, the angular streaking technique has the potential to be leveraged from the proof-of-principle stage to a robust and highly advanced diagnostic tool for all free-electron laser facilities, including high-repetition rate operation. In addition, these novel ML reconstruction procedures may also be used for better online X-ray pulse control and future FEL pulse shaping on demand.
	Further steps to a successful implementation of these advanced methods involve closing the gap between simulation and experimental data through an instrument-specific treatment of the measurement noise and a reliable concept for error and reliability estimation, which we will investigate in future work.

    \section*{Methods: Machine Learning Procedure Design}\label{sec:MLExpDesign}

    In real-world XFEL experiments, the spectrograms or pulse characteristics of individual SASE pulses have to be reconstructed from the detector image. 
    There are first approaches for deriving single-shot characteristics of rapid pulse sequences from high-repetition rate XFELs \cite{heider2019megahertz}. Unfortunately, they are only suitable to a limited degree for providing detailed insights via real-time processing during experimental campaigns. 
    
    \begin{figure}[!htbp]
	    \centering
	    \includegraphics[width=0.6\textwidth]{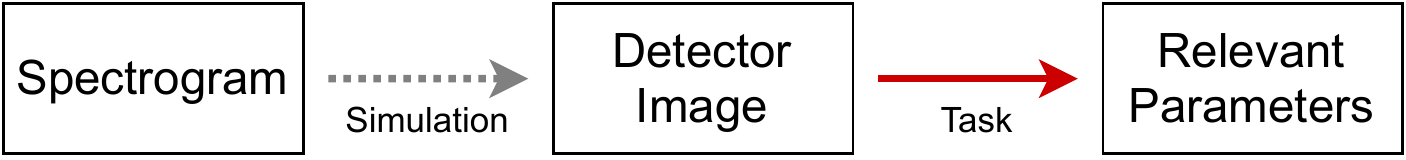}
	    \caption{We use a simulated spectrogram to construct according detector images. These images are used for training the NNs and to extract several different pulse characteristics afterward.}
	    \label{fig:sim_pipeline}
	\end{figure}
    
	Here, we apply specifically developed NNs on the angular streaking approach to demonstrate the possibility of a fast online pulse characterization, as NNs, particularly convolutional NNs, have proven to be suitable for similar challenges \cite{Ren2015, DroNet}.
	
	\subsection*{General Machine Learning  Problem Formulation} \label{subsec:general_ml_problem}
	
	For each pulse characteristic, we need to train a NN that takes detector images as inputs (cf.~Fig.~\ref{fig:sim_pipeline}). The outputs for each of the NNs vary and are listed below:
	
    \paragraph{\textbf{Kick}} The kick is the amplitude of the wave-like intensity distribution within the detector image (cf. Fig.~\ref{fig:spectrogram-detectorImage}). 
    When changing the kick, the spectrogram stays as is as the kick only affects the streaking signal captured in the detector image. That is the reason why the kick is easily extractable from the detector images. The NN has to solve a regression task, where the output is one number in the unit eV.
    
	\paragraph{\textbf{FWHM Pulse Duration}} The FWHM pulse duration is well extractable from the spectrogram, as it can be seen as $2.35 \cdot \sigma$ (cf. Eq.~\ref{eq:fwhm_duration}) in the direction of $x$ (time scale), with $\sigma$ being the standard deviation of the 2D gaussian distribution in x direction. The longer the FWHM pulse duration, the longer the distribution stretches in x direction. Within the detector image, a change of the pulse duration mostly affects the width and the peculiarity of the wave form. Here again, the NN has to solve a regression task, where the output is one number in the unit fs.
	
	\paragraph{\textbf{Auger Decay Time}} The Auger decay is visible in the spectrograms as well as the detector images. Within the spectrogram, the length of the tail after the 2D Gaussian distribution indicates the decay time. The longer the tail, the larger the decay. Within the detector image, a larger decay affects the distortion of the wave. Once more, the NN has to solve a regression task, where the output is one number in the unit fs.
	
	\paragraph{\textbf{Pulse Structure}} The pulse structure is the most challenging feature to extract, as the output itself consists of several values indicating the intensities of multiple spikes within the SASE pulses. By looking at the spectrogram in Fig.~\ref{fig:spectrogram-detectorImage} (c), one can see that the pulse structure can be derived by summing up the intensities at each point in time along the vertical axis. The pulse characteristic will be determined here as the intensity as a function of arrival time, where the intensity is integrated over all photon energies within the 6 eV spectral bandwidth. This leads to an output similar to Fig.~\ref{fig:pulseStructure}.
	In this case, the NN has to solve a regression task, where the output consists of several time steps in arbitrary intensity units.
	
	A note regarding the \textit{RMS Pulse Duration}: As the RMS pulse duration can be directly derived from the pulse structure, there is no need to train an independent NN for this pulse characteristic. 
	\mbox{}\\

	\noindent After the general examination of the ML problem, the next sections will look at how the ML pipeline looks in detail and how to successively address the individual ML problems above.
	
	\subsection*{Framework Conditions} \label{subsec:MLRequirements}
	
	In order to train NNs in a supervised manner, we require training data $\mathcal{D}_K$ of size $K \in \mathbb{N}$, which comprises $K$ simulated detector images 
	$\mathcal{X} = \{\mathbf{X}_i \in \mathbf{M}^{m \times n}(\mathbb{R}),\ i = 1, \dots, K \}$
	and $K$ corresponding pulse characteristics
	$\mathcal{L} = \{L_i \in \mathbb{R}^j_{+},\ i = 1, \dots, K \}$.
	Here, $m$ is the number of TOF detectors used within the complete spectrometer setup and $n$ displays the electron kinetic energy in intervals. 
	The size of $j$ changes according to the pulse characteristic that has to be predicted.
	In the following, we will refer to pulse characteristics as \textit{labels}. To verify the performance of a NN, we split $\mathcal{D}_K$ into two distinct sets, $\mathcal{D}_{\rm train}$ and $\mathcal{D}_{\rm test}$, such that $\mathcal{D}_K = \mathcal{D}_{\rm train} \cup \mathcal{D}_{\rm test}$. The accuracy of the NN is determined by $\mathcal{D}_{\rm test}$. 
	We train the NNs with $n$ distinct batches of detector images $\mathcal{B}_n \in \mathcal{D}_{\rm train}$. Similarly, we test the performance of the NNs with $m$ batches of detector images $\mathcal{B}_m \in \mathcal{D}_{\rm test}$. 
	To avoid overfitting of the NN, we utilize \textit{cross-validation}~\cite{BERRAR2019542}.
	
	Although we work in a simulation environment, it is reasonable to choose values that would correspond to real experimental data (cf. Fig.~\ref{fig:real_shot}). Therefore, we take previously acquired data from earlier experimental campaigns~\cite{hartmann2018attosecond} as an example. In particular this means:
	
	\begin{itemize}
	    \item For each of the two use cases, Ne 1s and KLL Auger electrons data, we generate a size of
	    $K = 4.4 \cdot 10^{6}$ samples to be predicted. Of these, $4 \cdot 10^{6}$ are used for training and $4 \cdot 10^{5}$ for testing.
	    \item Our angle-resolved spectrometer consists of $m = 16$ TOF detectors.
	    \item We fix the intervals in the TOF detectors to $n = 200$, with a varying energy bin size. 
	\end{itemize}
	
	More particularly, this means that the following NN architecture depends on the chosen parameters, though it can be adapted easily if, e.g., more TOF detectors are added.
	
	\subsection*{Preparing the Simulation Data} \label{subsec:MLDataPreparation}
	
	We derive artificial detector images for Ne 1s and KLL Auger electrons from the simulation environment as introduced by Hartmann et al. \cite{hartmann2018attosecond}. The kinetic energy of the 1s photoelectrons depends on the ionizing X-ray photon energy, which, in this case, is set to 1180 eV. We include a spectral bandwidth of 6 eV for the X-ray pulses, but omit the effect of a potential chirp in these simulations, which would only have a marginal effect on the parameters reconstructed in this study. A photon energy of 1180 eV results in Ne 1s photoelectron kinetic energies centered at $\sim$310 eV. The Auger electron kinetic energies are independent of the X-ray photon energy and bandwidth, with the main peak lying at $\sim$804 eV and a standard deviation determined by the detector resolution. The angular streaking maps the TOF measurements of 16 detectors distributed over 360 degree to a window of $35.3$ fs, as this is the duration of one optical cycle for the chosen streaking wavelength $\lambda=10.6\,\mathrm{\mu}$m of the circularly polarized laser in Hartmann \textit{et al.}\cite{hartmann2018attosecond}
	
	We chose the range and the precision based on the experimental implementation expected for real streaking measurements at XFELs. We want our models to estimate kicks in the range of $0-30$~eV, FWHM pulse durations in the range of $0.4-13.4$~fs, and decays in the range of $0-10$~fs. The temporal resolution of the pulse structure was equally chosen in accordance with the expected duration of the shortest features in the X-ray pulse intensity structures (the SASE spikes), leading to a grid size along the time axis for the pulse structure reconstruction of $441$~as.
	
	Figs.~\ref{fig:perfSpectrogram} and \ref{fig:perfDetectorImage} are simulated without artifacts. In \ref{fig:perfSpectrogram}, there is only one randomly placed Gaussian distribution present in the spectrogram. The underlying pulse structure is neglected so far.  
	To get closer to the real data (cf. Fig.~\ref{fig:real_shot}), we implement three steps. We add a pulse structure to the spectrogram and noise to the simulated detector images, and prepare the data for NN training by utilizing data normalization.

	\paragraph{\textbf{Step 1}: Adding a Pulse Structure to the Spectrogram}
	
	To achieve a SASE-like temporal structure in the spectrogram, we modulate the original Gaussian time
	distribution with a spiky intensity profile (cf. Fig.~\ref{fig:pulseStructure}). We obtain the latter by generating a comb of Gaussian spikes with randomized amplitudes and spike durations as predicted by theory for a typical setting of an XFEL in ultrashort-pulse mode \cite{Krinsky2003}.

	\paragraph{\textbf{Step 2}: Adding Noise to the Detector Image}
	
    Additional noise is added to $\mathbf{X} \in \mathcal{X}$, representing the intensity values for each pixel in the detector image, during training and testing as shown in Eq.~\ref{eq:noise}. A given percentage $p$ of the maximum intensity value $x_{\rm max}$ of $\mathbf{X} \in \mathcal{X}$ is used as an upper and lower bound of an equal distribution $\mathcal{G}$ to draw $w$ from:
        
    \begin{equation}
    \mathbf{X}_{\rm noisy} = \left( x_{i,j} + (x_{\rm max} \cdot w)\right)_{i=1,\dots,m, j=1,\dots,n}, w \sim \mathcal{G}(-p,p),
    \label{eq:noise}
    \end{equation}
        
    \noindent Figure~\ref{fig:noiseLevels} displays a detector image with added noise ($p = 0.3$).

    \paragraph{\textbf{Step 3}: Normalizing the Data}
    
    It is evident, that the range of the intensity values differs from case to case.
    To counteract this, we perform a min-max-normalization for each $\mathbf{X} \in \mathcal{X}$.
    Therefore, the minimum ($x_{\rm min}$) and maximum ($x_{\rm max}$) intensity value of $\mathbf{X}$ are used to perform the transformation for each pixel value $x_{k,l}$:

    \begin{equation}
    \mathbf{X}_{\rm norm} = \left( \frac{x_{k,l} - x_{\rm min}}{x_{\rm max}-x_{\rm min}} \right)_{k=1,\dots,m, l=1,\dots,n},
    \end{equation}
    
    \noindent After normalization, all values in $\mathbf{X}_{\rm norm}$ lie within the interval $[0,1]$.
	
	\subsection*{Designing the Machine Learning Models} \label{subsec:MLModelDesign}
	
	As we want to extract information from images, the most intuitive solution is to use a convolutional NN~\cite{krizhevsky2012imagenet,Gu2018}, which uses convolutions and pooling to extract low- and high-level features such as edges and predicts an estimate of those features using fully-connected layers. In our case, the estimates ideally should correspond to the target pulse characteristics.
	
	\subsubsection*{Architecture}
	
\begin{figure*}[!ht]
	    \vspace*{-2em}
	    \centering
	    \includegraphics[width=0.8\textwidth]{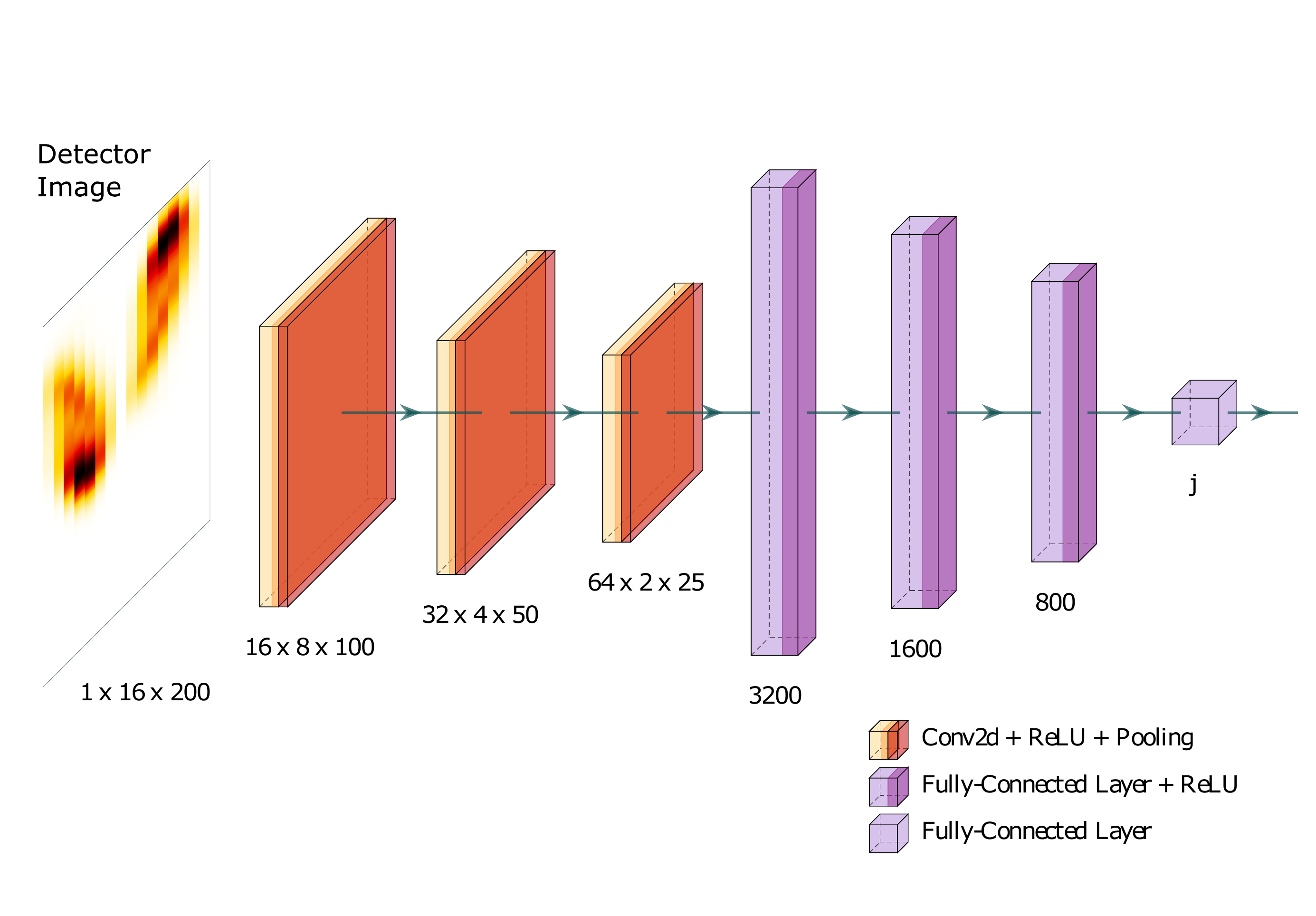}
	    \caption{The convolutional neural network architecture used in the present approach. The dimensions of the first layers (orange) are displayed in [channel, height, width]. The dimensions of the fully-connected layers (violett) show the number of used neurons.}
	    \label{fig:nn_architecture}
	\end{figure*}

	One key problem regarding the choice of a proper NN architecture is the dimensionality of the detector images, which are not equal in size and therefore not symmetrical. This fact needed to be taken into account in the design of the NN. Furthermore, we wanted our NN architecture to be as dense as possible to ensure generalization and performance for reliable online operation. After testing several architecture configurations with different numbers of layers and neurons, the most suitable network architecture for our problem is an NN with three convolutional blocks (cf. Fig.~\ref{fig:nn_architecture}). Each block contains a convolutional layer, followed by an activation function and a max-pooling layer. The convolutional layers use a  $3x3$ kernel, stride of $1x1$ and $1x1$ zero-padding. The pooling layers use a $3x3$ kernel, stride of $2x2$ and $1x1$ zero-padding. 
	Architectures with less than three convolutional blocks could not grasp all required features necessary to derive the underlying mapping from detector image to the desired label.
	The NN is specifically designed to cut both dimensions, i.e., width and height of the image, in half after each block. A filter size of $[16, 32, 64]$ for the respective convolutional layers has proven to be sufficient. For the fully-connected stage (except the last layer), we use three layers with $[3200, 1600, 800]$ neurons, respectively. Architectures with less than three layers within the fully-connected part did not transmit enough information to be able to solve the problem decently. The size of the last layer depends on the label to predict, i.e., the size of $j$ in $\mathcal{L}$.
	
	When predicting the \textit{kick}, \textit{FWHM pulse duration}, or \textit{decay time}, $j = 1$. When predicting the \textit{pulse structure}, $j$ corresponds to the dimension of the spectrogram's x-axis. We utilize the mean squared error loss function to train and optimize the network as the prediction of the pulse characteristics is a regression task in all cases.
	
	\subsubsection*{Hyperparameter Optimization}
	
	The NN architecture is not the only choice to be considered. Especially when training the NNs, appropriately chosen hyperparameters are important to achieve efficient and goal-oriented training. Important parameters in this context are the \textit{batch size}, \textit{type of activation function}, \textit{optimizer}, and \textit{learning rate}. To find the best suitable combination of hyperparameters, we performed a grid-search on distinct data sets using the approach from before with the following values:
	
	\begin{itemize}
	    \item \textit{Batch size}: [64, 128, 256, 512, 1024]
	    \item \textit{Activation function}: [ReLU, Sigmoid]
	    \item \textit{Optimizer}: [Adam, SGD (with Momentum)]
	    \item \textit{Learning rate}: [0.01, 0.001, 0.0001, 0.00001]
	\end{itemize}
	
	After NN training, we evaluated the respective parameter combinations according to the following criteria:
	
	\begin{itemize}
	    \item \textit{Criterion 1}: The test loss (after inputting $\mathcal{D}_{\rm test}$ into the trained NN) should be minimal.
	    \item \textit{Criterion 2}: The standard deviation of the test loss curve should be minimal to penalize slow convergence and overfitting. 
	\end{itemize}
	
	In general, it should be noted that there is not only one combination of hyperparameters that achieves good results during training. Nevertheless, there has been an evident leader. The best hyperparameter configuration for all labels is a batch size of $64$, a ReLU activation function, a learning rate of $0.0001$, and Adam as optimizer.

\bibliography{references}

\begin{thebibliography}{10}
\urlstyle{rm}
\expandafter\ifx\csname url\endcsname\relax
  \def\url#1{\texttt{#1}}\fi
\expandafter\ifx\csname urlprefix\endcsname\relax\def\urlprefix{URL }\fi
\expandafter\ifx\csname doiprefix\endcsname\relax\def\doiprefix{DOI: }\fi
\providecommand{\bibinfo}[2]{#2}
\providecommand{\eprint}[2][]{\url{#2}}

\bibitem{mazza2020mapping}
\bibinfo{author}{Mazza, T.} \emph{et~al.}
\newblock \bibinfo{journal}{\bibinfo{title}{{Mapping Resonance Structures in
  Transient Core-Ionized Atoms}}}.
\newblock {\emph{\JournalTitle{Physical Review X}}}
  \textbf{\bibinfo{volume}{10}}, \bibinfo{pages}{041056}
  (\bibinfo{year}{2020}).

\bibitem{SingleMolImTrans}
\bibinfo{author}{Ho, P.~J.} \emph{et~al.}
\newblock \bibinfo{journal}{\bibinfo{title}{The role of transient resonances
  for ultra-fast imaging of single sucrose nanoclusters}}.
\newblock {\emph{\JournalTitle{Nature communications}}}
  \textbf{\bibinfo{volume}{11}}, \bibinfo{pages}{1--9} (\bibinfo{year}{2020}).

\bibitem{Roadmap}
\bibinfo{author}{Young, L.} \emph{et~al.}
\newblock \bibinfo{journal}{\bibinfo{title}{Roadmap of ultrafast x-ray atomic
  and molecular physics}}.
\newblock {\emph{\JournalTitle{Journal of Physics B: Atomic, Molecular and
  Optical Physics}}} \textbf{\bibinfo{volume}{51}}, \bibinfo{pages}{032003}
  (\bibinfo{year}{2018}).

\bibitem{eichmann2020photon}
\bibinfo{author}{Eichmann, U.} \emph{et~al.}
\newblock \bibinfo{journal}{\bibinfo{title}{{Photon-recoil imaging: Expanding
  the view of nonlinear x-ray physics}}}.
\newblock {\emph{\JournalTitle{Science}}} \textbf{\bibinfo{volume}{369}},
  \bibinfo{pages}{1630--1633} (\bibinfo{year}{2020}).

\bibitem{decking2020mhz}
\bibinfo{author}{Decking, W.} \emph{et~al.}
\newblock \bibinfo{journal}{\bibinfo{title}{{A MHz-repetition-rate hard X-ray
  free-electron laser driven by a superconducting linear accelerator}}}.
\newblock {\emph{\JournalTitle{Nature Photonics}}}
  \textbf{\bibinfo{volume}{14}}, \bibinfo{pages}{391--397}
  (\bibinfo{year}{2020}).

\bibitem{milton2001exponential}
\bibinfo{author}{Milton, S.} \emph{et~al.}
\newblock \bibinfo{journal}{\bibinfo{title}{Exponential gain and saturation of
  a self-amplified spontaneous emission free-electron laser}}.
\newblock {\emph{\JournalTitle{Science}}} \textbf{\bibinfo{volume}{292}},
  \bibinfo{pages}{2037--2041} (\bibinfo{year}{2001}).

\bibitem{Tzallas2003}
\bibinfo{author}{Tzallas, P.}, \bibinfo{author}{Charalambidis, D.},
  \bibinfo{author}{Papadogiannis, N.~A.}, \bibinfo{author}{Witte, K.} \&
  \bibinfo{author}{Tsakiris, G.~D.}
\newblock \bibinfo{journal}{\bibinfo{title}{{Direct observation of attosecond
  light bunching}}}.
\newblock {\emph{\JournalTitle{Nature}}} \textbf{\bibinfo{volume}{426}},
  \bibinfo{pages}{267--271} (\bibinfo{year}{2003}).

\bibitem{Paul2001}
\bibinfo{author}{Paul, P.~M.} \emph{et~al.}
\newblock \bibinfo{journal}{\bibinfo{title}{{Observation of a train of
  attosecond pulses from high harmonic generation.}}}
\newblock {\emph{\JournalTitle{Science}}} \textbf{\bibinfo{volume}{292}},
  \bibinfo{pages}{1689--92} (\bibinfo{year}{2001}).

\bibitem{Kienberger2004}
\bibinfo{author}{Kienberger, R.} \emph{et~al.}
\newblock \bibinfo{journal}{\bibinfo{title}{{Atomic transient recorder}}}.
\newblock {\emph{\JournalTitle{Nature}}} \textbf{\bibinfo{volume}{427}},
  \bibinfo{pages}{817--821} (\bibinfo{year}{2004}).

\bibitem{serkez2020opportunities}
\bibinfo{author}{Serkez, S.} \emph{et~al.}
\newblock \bibinfo{journal}{\bibinfo{title}{{Opportunities for two-color
  experiments in the soft X-ray regime at the european XFEL}}}.
\newblock {\emph{\JournalTitle{Applied Sciences}}}
  \textbf{\bibinfo{volume}{10}}, \bibinfo{pages}{2728} (\bibinfo{year}{2020}).

\bibitem{behrens2014few}
\bibinfo{author}{Behrens, C.} \emph{et~al.}
\newblock \bibinfo{journal}{\bibinfo{title}{{Few-femtosecond time-resolved
  measurements of X-ray free-electron lasers}}}.
\newblock {\emph{\JournalTitle{Nature communications}}}
  \textbf{\bibinfo{volume}{5}}, \bibinfo{pages}{1--7} (\bibinfo{year}{2014}).

\bibitem{hartmann2018attosecond}
\bibinfo{author}{Hartmann, N.} \emph{et~al.}
\newblock \bibinfo{journal}{\bibinfo{title}{{Attosecond time--energy structure
  of X-ray free-electron laser pulses}}}.
\newblock {\emph{\JournalTitle{Nature Photonics}}}
  \textbf{\bibinfo{volume}{12}}, \bibinfo{pages}{215} (\bibinfo{year}{2018}).

\bibitem{duris2020tunable}
\bibinfo{author}{Duris, J.} \emph{et~al.}
\newblock \bibinfo{journal}{\bibinfo{title}{{Tunable isolated attosecond X-ray
  pulses with gigawatt peak power from a free-electron laser}}}.
\newblock {\emph{\JournalTitle{Nature Photonics}}}
  \textbf{\bibinfo{volume}{14}}, \bibinfo{pages}{30--36}
  (\bibinfo{year}{2020}).

\bibitem{driver2020attosecond}
\bibinfo{author}{Driver, T.} \emph{et~al.}
\newblock \bibinfo{journal}{\bibinfo{title}{Attosecond transient absorption
  spooktroscopy: a ghost imaging approach to ultrafast absorption
  spectroscopy}}.
\newblock {\emph{\JournalTitle{Physical Chemistry Chemical Physics}}}
  \textbf{\bibinfo{volume}{22}}, \bibinfo{pages}{2704--2712}
  (\bibinfo{year}{2020}).

\bibitem{Bonifacio1994}
\bibinfo{author}{Bonifacio, R.}, \bibinfo{author}{De~Salvo, L.},
  \bibinfo{author}{Pierini, P.}, \bibinfo{author}{Piovella, N.} \&
  \bibinfo{author}{Pellegrini, C.}
\newblock \bibinfo{journal}{\bibinfo{title}{Spectrum, temporal structure, and
  fluctuations in a high-gain free-electron laser starting from noise}}.
\newblock {\emph{\JournalTitle{Phys. Rev. Lett.}}}
  \textbf{\bibinfo{volume}{73}}, \bibinfo{pages}{70--73}
  (\bibinfo{year}{1994}).

\bibitem{Neutze2000}
\bibinfo{author}{Neutze, R.}, \bibinfo{author}{Wouts, R.},
  \bibinfo{author}{van~der Spoel, D.}, \bibinfo{author}{Weckert, E.} \&
  \bibinfo{author}{Hajdu, J.}
\newblock \bibinfo{journal}{\bibinfo{title}{{Potential for biomolecular imaging
  with femtosecond X-ray pulses}}}.
\newblock {\emph{\JournalTitle{Nature}}} \textbf{\bibinfo{volume}{406}},
  \bibinfo{pages}{752--7} (\bibinfo{year}{2000}).

\bibitem{Barty2008}
\bibinfo{author}{Barty, A.} \emph{et~al.}
\newblock \bibinfo{journal}{\bibinfo{title}{{Ultrafast single-shot diffraction
  imaging of nanoscale dynamics}}}.
\newblock {\emph{\JournalTitle{Nature Photonics}}}
  \textbf{\bibinfo{volume}{2}}, \bibinfo{pages}{415--419}
  (\bibinfo{year}{2008}).

\bibitem{Young2010}
\bibinfo{author}{Young, L.} \emph{et~al.}
\newblock \bibinfo{journal}{\bibinfo{title}{{Femtosecond electronic response of
  atoms to ultra-intense X-rays}}}.
\newblock {\emph{\JournalTitle{Nature}}} \textbf{\bibinfo{volume}{466}},
  \bibinfo{pages}{56--61} (\bibinfo{year}{2010}).

\bibitem{Rudenko2017}
\bibinfo{author}{Rudenko, A.} \emph{et~al.}
\newblock \bibinfo{journal}{\bibinfo{title}{{Femtosecond response of polyatomic
  molecules to ultra-intense hard X-rays}}}.
\newblock {\emph{\JournalTitle{Nature}}} \textbf{\bibinfo{volume}{546}},
  \bibinfo{pages}{129--132} (\bibinfo{year}{2017}).

\bibitem{eckle2008attosecond}
\bibinfo{author}{Eckle, P.} \emph{et~al.}
\newblock \bibinfo{journal}{\bibinfo{title}{Attosecond angular streaking}}.
\newblock {\emph{\JournalTitle{Nature Physics}}} \textbf{\bibinfo{volume}{4}},
  \bibinfo{pages}{565--570} (\bibinfo{year}{2008}).

\bibitem{Constant1997}
\bibinfo{author}{Constant, E.}, \bibinfo{author}{Taranukhin, V.},
  \bibinfo{author}{Stolow, A.} \& \bibinfo{author}{Corkum, P.}
\newblock \bibinfo{journal}{\bibinfo{title}{{Methods for the measurement of the
  duration of high-harmonic pulses}}}.
\newblock {\emph{\JournalTitle{Physical Review A}}}
  \textbf{\bibinfo{volume}{56}}, \bibinfo{pages}{3870--3878}
  (\bibinfo{year}{1997}).

\bibitem{Itatani2002}
\bibinfo{author}{Itatani, J.} \emph{et~al.}
\newblock \bibinfo{journal}{\bibinfo{title}{{Attosecond Streak Camera}}}.
\newblock {\emph{\JournalTitle{Physical Review Letters}}}
  \textbf{\bibinfo{volume}{88}}, \bibinfo{pages}{173903}
  (\bibinfo{year}{2002}).

\bibitem{Kazansky2016}
\bibinfo{author}{Kazansky, A.~K.}, \bibinfo{author}{Bozhevolnov, A.~V.},
  \bibinfo{author}{Sazhina, I.~P.} \& \bibinfo{author}{Kabachnik, N.~M.}
\newblock \bibinfo{journal}{\bibinfo{title}{{Interference effects in angular
  streaking with a rotating terahertz field}}}.
\newblock {\emph{\JournalTitle{Physical Review A}}}
  \textbf{\bibinfo{volume}{93}}, \bibinfo{pages}{013407}
  (\bibinfo{year}{2016}).

\bibitem{Bionta2011}
\bibinfo{author}{Bionta, M.~R.} \emph{et~al.}
\newblock \bibinfo{journal}{\bibinfo{title}{{Spectral encoding of x-ray/optical
  relative delay}}}.
\newblock {\emph{\JournalTitle{Optics Express}}} \textbf{\bibinfo{volume}{19}},
  \bibinfo{pages}{21855--65} (\bibinfo{year}{2011}).

\bibitem{Hartmann2014}
\bibinfo{author}{Hartmann, N.} \emph{et~al.}
\newblock \bibinfo{journal}{\bibinfo{title}{{Sub-femtosecond precision
  measurement of relative X-ray arrival time for free-electron lasers}}}.
\newblock {\emph{\JournalTitle{Nature Photonics}}}
  \textbf{\bibinfo{volume}{8}}, \bibinfo{pages}{706--709}
  (\bibinfo{year}{2014}).

\bibitem{Diez2021}
\bibinfo{author}{Diez, M.} \emph{et~al.}
\newblock \bibinfo{journal}{\bibinfo{title}{{A self-referenced in-situ arrival
  time monitor for X-ray free-electron lasers}}}.
\newblock {\emph{\JournalTitle{Scientific Reports}}}
  \textbf{\bibinfo{volume}{11}}, \bibinfo{pages}{3562} (\bibinfo{year}{2021}).

\bibitem{Fuji2005}
\bibinfo{author}{Fuji, T.} \emph{et~al.}
\newblock \bibinfo{journal}{\bibinfo{title}{{Monolithic carrier-envelope
  phase-stabilization scheme}}}.
\newblock {\emph{\JournalTitle{Optics Letters}}} \textbf{\bibinfo{volume}{30}},
  \bibinfo{pages}{332} (\bibinfo{year}{2005}).

\bibitem{ocelot}
\bibinfo{author}{Agapov, I.}, \bibinfo{author}{Geloni, G.},
  \bibinfo{author}{Tomin, S.} \& \bibinfo{author}{Zagorodnov, I.}
\newblock \bibinfo{journal}{\bibinfo{title}{{OCELOT: A software framework for
  synchrotron light source and FEL studies}}}.
\newblock {\emph{\JournalTitle{Nuclear Instruments and Methods in Physics
  Research Section A: Accelerators, Spectrometers, Detectors and Associated
  Equipment}}} \textbf{\bibinfo{volume}{768}}, \bibinfo{pages}{151--156}
  (\bibinfo{year}{2014}).

\bibitem{rudek2012ultra}
\bibinfo{author}{Rudek, B.} \emph{et~al.}
\newblock \bibinfo{journal}{\bibinfo{title}{{Ultra-efficient ionization of
  heavy atoms by intense X-ray free-electron laser pulses}}}.
\newblock {\emph{\JournalTitle{Nature photonics}}}
  \textbf{\bibinfo{volume}{6}}, \bibinfo{pages}{858--865}
  (\bibinfo{year}{2012}).

\bibitem{Picon2016}
\bibinfo{author}{Pic{\'{o}}n, A.} \emph{et~al.}
\newblock \bibinfo{journal}{\bibinfo{title}{{Hetero-site-specific X-ray
  pump-probe spectroscopy for femtosecond intramolecular dynamics}}}.
\newblock {\emph{\JournalTitle{Nature Communications}}}
  \textbf{\bibinfo{volume}{7}}, \bibinfo{pages}{11652} (\bibinfo{year}{2016}).

\bibitem{li2021electron}
\bibinfo{author}{Li, X.} \emph{et~al.}
\newblock \bibinfo{journal}{\bibinfo{title}{{Electron-ion coincidence
  measurements of molecular dynamics with intense X-ray pulses}}}.
\newblock {\emph{\JournalTitle{Scientific reports}}}
  \textbf{\bibinfo{volume}{11}}, \bibinfo{pages}{1--12} (\bibinfo{year}{2021}).

\bibitem{sorokin2000measurement}
\bibinfo{author}{Sorokin, E.}, \bibinfo{author}{Tempea, G.} \&
  \bibinfo{author}{Brabec, T.}
\newblock \bibinfo{journal}{\bibinfo{title}{Measurement of the root-mean-square
  width and the root-mean-square chirp in ultrafast optics}}.
\newblock {\emph{\JournalTitle{JOSA B}}} \textbf{\bibinfo{volume}{17}},
  \bibinfo{pages}{146--150} (\bibinfo{year}{2000}).

\bibitem{Krinsky2003}
\bibinfo{author}{Krinsky, S.} \& \bibinfo{author}{Gluckstern, R.}
\newblock \bibinfo{journal}{\bibinfo{title}{{Analysis of statistical
  correlations and intensity spiking in the self-amplified spontaneous-emission
  free-electron laser}}}.
\newblock {\emph{\JournalTitle{Physical Review Special Topics - Accelerators
  and Beams}}} \textbf{\bibinfo{volume}{6}}, \bibinfo{pages}{050701}
  (\bibinfo{year}{2003}).

\bibitem{Haynes2021}
\bibinfo{author}{Haynes, D.~C.} \emph{et~al.}
\newblock \bibinfo{journal}{\bibinfo{title}{{Clocking Auger electrons}}}.
\newblock {\emph{\JournalTitle{Nature Physics}}} \textbf{\bibinfo{volume}{17}},
  \bibinfo{pages}{512--518} (\bibinfo{year}{2021}).

\bibitem{lederer}
\bibinfo{author}{Pergament, M.} \emph{et~al.}
\newblock \bibinfo{journal}{\bibinfo{title}{{Versatile optical laser system for
  experiments at the European X-ray free-electron laser facility}}}.
\newblock {\emph{\JournalTitle{Opt. Express}}} \textbf{\bibinfo{volume}{24}},
  \bibinfo{pages}{29349--29359} (\bibinfo{year}{2016}).

\bibitem{Gal2016}
\bibinfo{author}{Gal, Y.} \& \bibinfo{author}{Ghahramani, Z.}
\newblock \bibinfo{title}{{Dropout as a Bayesian Approximation: Representing
  Model Uncertainty in Deep Learning}}.
\newblock In \bibinfo{editor}{Balcan, M.~F.} \& \bibinfo{editor}{Weinberger,
  K.~Q.} (eds.) \emph{\bibinfo{booktitle}{Proceedings of The 33rd International
  Conference on Machine Learning}}, vol.~\bibinfo{volume}{48} of
  \emph{\bibinfo{series}{Proceedings of Machine Learning Research}},
  \bibinfo{pages}{1050--1059} (\bibinfo{publisher}{PMLR}, \bibinfo{address}{New
  York, New York, USA}, \bibinfo{year}{2016}).

\bibitem{Teye2018}
\bibinfo{author}{Teye, M.}, \bibinfo{author}{Azizpour, H.} \&
  \bibinfo{author}{Smith, K.}
\newblock \bibinfo{title}{{B}ayesian uncertainty estimation for batch
  normalized deep networks}.
\newblock In \bibinfo{editor}{Dy, J.} \& \bibinfo{editor}{Krause, A.} (eds.)
  \emph{\bibinfo{booktitle}{Proceedings of the 35th International Conference on
  Machine Learning}}, vol.~\bibinfo{volume}{80} of
  \emph{\bibinfo{series}{Proceedings of Machine Learning Research}},
  \bibinfo{pages}{4907--4916} (\bibinfo{publisher}{PMLR},
  \bibinfo{year}{2018}).

\bibitem{kendall2017uncertainties}
\bibinfo{author}{Kendall, A.} \& \bibinfo{author}{Gal, Y.}
\newblock \bibinfo{title}{{What uncertainties do we need in Bayesian deep
  learning for computer vision?}}
\newblock In \emph{\bibinfo{booktitle}{Proceedings of the 31st International
  Conference on Neural Information Processing Systems}},
  \bibinfo{pages}{5580--5590} (\bibinfo{year}{2017}).

\bibitem{Vincent2010}
\bibinfo{author}{Vincent, P.}, \bibinfo{author}{Larochelle, H.},
  \bibinfo{author}{Lajoie, I.}, \bibinfo{author}{Bengio, Y.} \&
  \bibinfo{author}{Manzagol, P.-A.}
\newblock \bibinfo{journal}{\bibinfo{title}{{Stacked Denoising Autoencoders:
  Learning Useful Representations in a Deep Network with a Local Denoising
  Criterion}}}.
\newblock {\emph{\JournalTitle{Journal of Machine Learning Research}}}
  \textbf{\bibinfo{volume}{11}}, \bibinfo{pages}{3371--3408}
  (\bibinfo{year}{2010}).

\bibitem{Fontenla-Romero2013}
\bibinfo{author}{Fontenla-Romero, {\'{O}}.}, \bibinfo{author}{Martinez-Rego,
  D.}, \bibinfo{author}{Guijarro-Berdi{\~{n}}as, B.},
  \bibinfo{author}{P{\'{e}}rez-S{\'{a}}nchez, B.} \&
  \bibinfo{author}{Peteiro-Barral, D.}
\newblock \bibinfo{title}{{Online machine learning}}.
\newblock In \emph{\bibinfo{booktitle}{Efficiency and Scalability Methods for
  Computational Intellect}}, \bibinfo{pages}{27--54} (\bibinfo{publisher}{IGI
  Global}, \bibinfo{year}{2013}).

\bibitem{Kirkpatrick3521}
\bibinfo{author}{Kirkpatrick, J.} \emph{et~al.}
\newblock \bibinfo{journal}{\bibinfo{title}{Overcoming catastrophic forgetting
  in neural networks}}.
\newblock {\emph{\JournalTitle{Proceedings of the National Academy of
  Sciences}}} \textbf{\bibinfo{volume}{114}}, \bibinfo{pages}{3521--3526}
  (\bibinfo{year}{2017}).

\bibitem{PARISI201954}
\bibinfo{author}{Parisi, G.~I.}, \bibinfo{author}{Kemker, R.},
  \bibinfo{author}{Part, J.~L.}, \bibinfo{author}{Kanan, C.} \&
  \bibinfo{author}{Wermter, S.}
\newblock \bibinfo{journal}{\bibinfo{title}{{Continual lifelong learning with
  neural networks: A review}}}.
\newblock {\emph{\JournalTitle{Neural Networks}}}
  \textbf{\bibinfo{volume}{113}}, \bibinfo{pages}{54--71}
  (\bibinfo{year}{2019}).

\bibitem{He2021}
\bibinfo{author}{He, Y.} \& \bibinfo{author}{Sick, B.}
\newblock \bibinfo{journal}{\bibinfo{title}{{CLeaR: An adaptive continual
  learning framework for regression tasks}}}.
\newblock {\emph{\JournalTitle{AI Perspectives 2021 3:1}}}
  \textbf{\bibinfo{volume}{3}}, \bibinfo{pages}{1--16} (\bibinfo{year}{2021}).

\bibitem{heider2019megahertz}
\bibinfo{author}{Heider, R.} \emph{et~al.}
\newblock \bibinfo{journal}{\bibinfo{title}{Megahertz-compatible angular
  streaking with few-femtosecond resolution at x-ray free-electron lasers}}.
\newblock {\emph{\JournalTitle{Phys. Rev. A}}} \textbf{\bibinfo{volume}{100}},
  \bibinfo{pages}{053420} (\bibinfo{year}{2019}).

\bibitem{Ren2015}
\bibinfo{author}{Ren, S.}, \bibinfo{author}{He, K.}, \bibinfo{author}{Girshick,
  R.} \& \bibinfo{author}{Sun, J.}
\newblock \bibinfo{title}{Faster r-cnn: Towards real-time object detection with
  region proposal networks}.
\newblock In \bibinfo{editor}{Cortes, C.}, \bibinfo{editor}{Lawrence, N.},
  \bibinfo{editor}{Lee, D.}, \bibinfo{editor}{Sugiyama, M.} \&
  \bibinfo{editor}{Garnett, R.} (eds.) \emph{\bibinfo{booktitle}{Advances in
  Neural Information Processing Systems}}, vol.~\bibinfo{volume}{28}
  (\bibinfo{publisher}{Curran Associates, Inc.}, \bibinfo{year}{2015}).

\bibitem{DroNet}
\bibinfo{author}{Kyrkou, C.}, \bibinfo{author}{Plastiras, G.},
  \bibinfo{author}{Theocharides, T.}, \bibinfo{author}{Venieris, S.~I.} \&
  \bibinfo{author}{Bouganis, C.-S.}
\newblock \bibinfo{title}{{DroNet: Efficient convolutional neural network
  detector for real-time UAV applications}}.
\newblock In \emph{\bibinfo{booktitle}{2018 Design, Automation Test in Europe
  Conference Exhibition}}, \bibinfo{pages}{967--972} (\bibinfo{year}{2018}).

\bibitem{BERRAR2019542}
\bibinfo{author}{Berrar, D.}
\newblock \bibinfo{title}{{Cross-Validation}}.
\newblock In \bibinfo{editor}{Ranganathan, S.}, \bibinfo{editor}{Gribskov, M.},
  \bibinfo{editor}{Nakai, K.} \& \bibinfo{editor}{Schönbach, C.} (eds.)
  \emph{\bibinfo{booktitle}{Encyclopedia of Bioinformatics and Computational
  Biology}}, \bibinfo{pages}{542--545} (\bibinfo{publisher}{Academic Press},
  \bibinfo{address}{Oxford}, \bibinfo{year}{2019}).

\bibitem{krizhevsky2012imagenet}
\bibinfo{author}{Krizhevsky, A.}, \bibinfo{author}{Sutskever, I.} \&
  \bibinfo{author}{Hinton, G.~E.}
\newblock \bibinfo{journal}{\bibinfo{title}{Imagenet classification with deep
  convolutional neural networks}}.
\newblock {\emph{\JournalTitle{Advances in neural information processing
  systems}}} \textbf{\bibinfo{volume}{25}}, \bibinfo{pages}{1097--1105}
  (\bibinfo{year}{2012}).

\bibitem{Gu2018}
\bibinfo{author}{Gu, J.} \emph{et~al.}
\newblock \bibinfo{journal}{\bibinfo{title}{{Recent advances in convolutional
  neural networks}}}.
\newblock {\emph{\JournalTitle{Pattern Recognition}}}
  \textbf{\bibinfo{volume}{77}}, \bibinfo{pages}{354--377}
  (\bibinfo{year}{2018}).

\end{thebibliography}

\section*{Acknowledgements}

This research was supported by the project SpeAR\_XFEL (05K2019) funded by the BMBF (German Federal Ministry of Education and Research). We gratefully acknowledge the assistance and support of the Joint Laboratory Artificial Intelligence Methods for Experiment Design (AIM-ED) between Helmholtzzentrum für Materialien und Energie, Berlin, and the University of Kassel. MI acknowledges funding by the VW foundation for a Peter-Paul-Ewald Fellowship.

This version of the article has been accepted for publication, after peer review but is not the Version of Record and does not reflect post-acceptance improvements, or any corrections. The Version of Record is available online at:\\ \url{https://doi.org/10.1038/s41598-022-21646-x}

\section*{Author contributions statement}


\textbf{Kristina Dingel}: Conceptualization, Methodology, Software, Investigation, Writing - Original Draft, Visualization.

\noindent \textbf{Thorsten Otto}: Conceptualization, Methodology, Software, Investigation, Writing - Original Draft, Visualization.

\noindent \textbf{Lutz Marder}: Methodology, Software, Investigation, Writing - Original Draft, Visualization.

\noindent \textbf{Lars Funke}: Writing - Review \& Editing, Visualization.

\noindent \textbf{Arne Held}: Writing - Review \& Editing.

\noindent \textbf{Sara Savio}: Writing - Review \& Editing.

\noindent \textbf{Andreas Hans}: Writing - Review \& Editing.

\noindent \textbf{Gregor Hartmann}: Software, Writing - Review \& Editing.

\noindent \textbf{David Meier}: Writing - Review \& Editing.

\noindent \textbf{Jens Viefhaus}: Writing - Review \& Editing.

\noindent \textbf{Bernhard Sick}: Writing - Review \& Editing, Supervision, Funding acquisition.

\noindent \textbf{Arno Ehresmann}: Writing - Review \& Editing.

\noindent \textbf{Markus Ilchen}: Conceptualization, Writing - Original Draft, Supervision.

\noindent \textbf{Wolfram Helml}: Conceptualization, Writing - Original Draft, Supervision, Project administration, Funding acquisition.

\section*{Additional information}

\textbf{Data and Code Availability} A repository containing the detector image analysis software and according simulation data will be provided on request;

\noindent \textbf{Competing interests} The authors declare no competing interests.

\end{document}


%
%


\section*{Supplementary Information}

\renewcommand{\figurename}{Supplementary Figure}
\setcounter{figure}{0} 

\begin{figure}[!htbp]
    \centering
    \includegraphics[width=.6\textwidth]{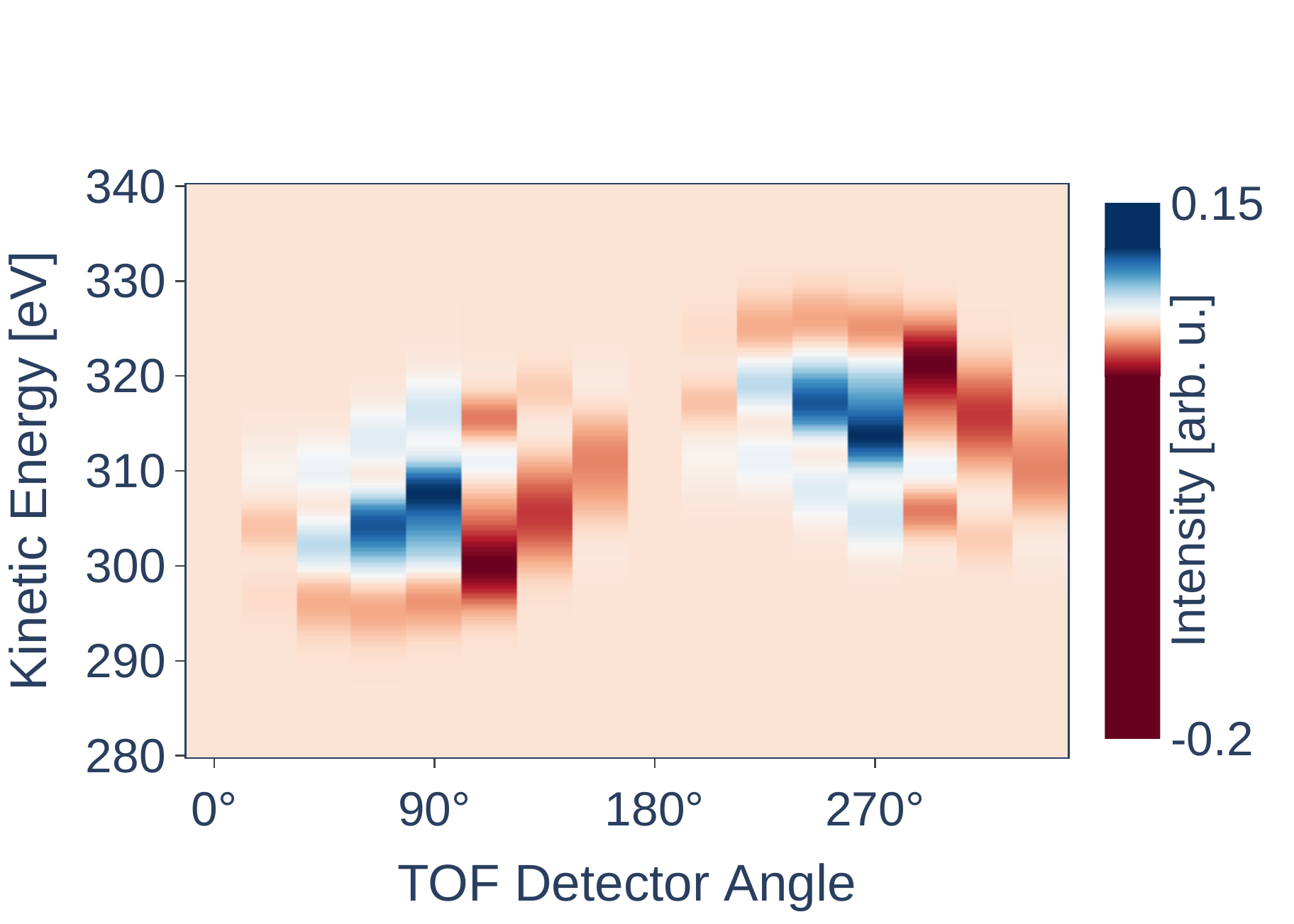}
    \caption{Difference between detector images in Fig.~2 (f) and (g). On closer inspection, it is evident that the intensities shift after adding a pulse structure. }
    \label{fig:detector_diff}
\end{figure}

\begin{figure}[!htbp]
    \centering
    \subfloat[NN estimate ($4.8$ fs) of the FWHM pulse duration target ($4.9$ fs) with a large kick ($22.5$ eV).]{\includegraphics[width=0.49\textwidth]{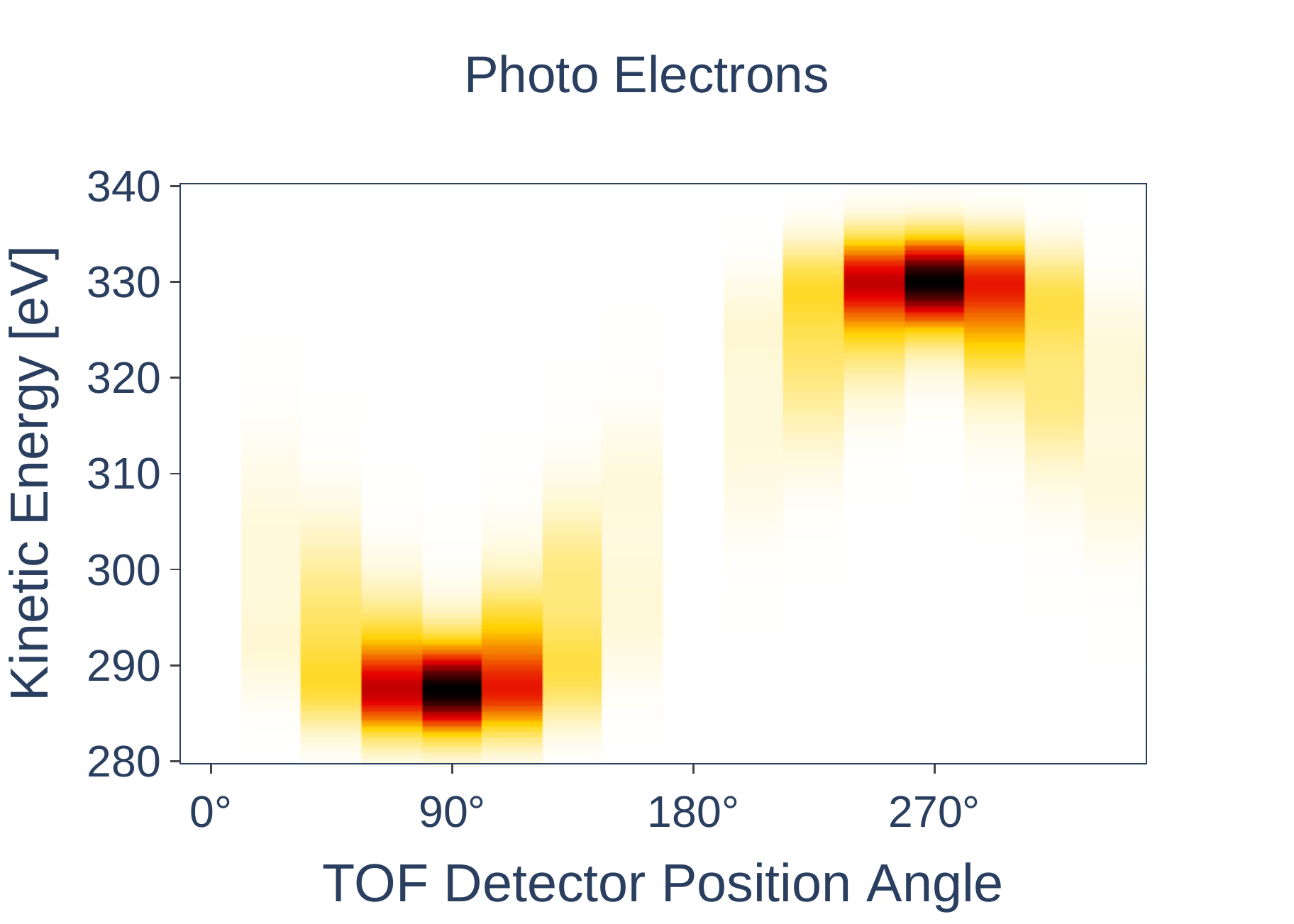}\label{fig:5a}}
    \hspace{1em}
    \subfloat[NN estimate ($6.3$ fs) of the FWHM pulse duration target ($4.9$ fs) with a small kick ($3.0$ eV).]{\includegraphics[width=0.49\textwidth]{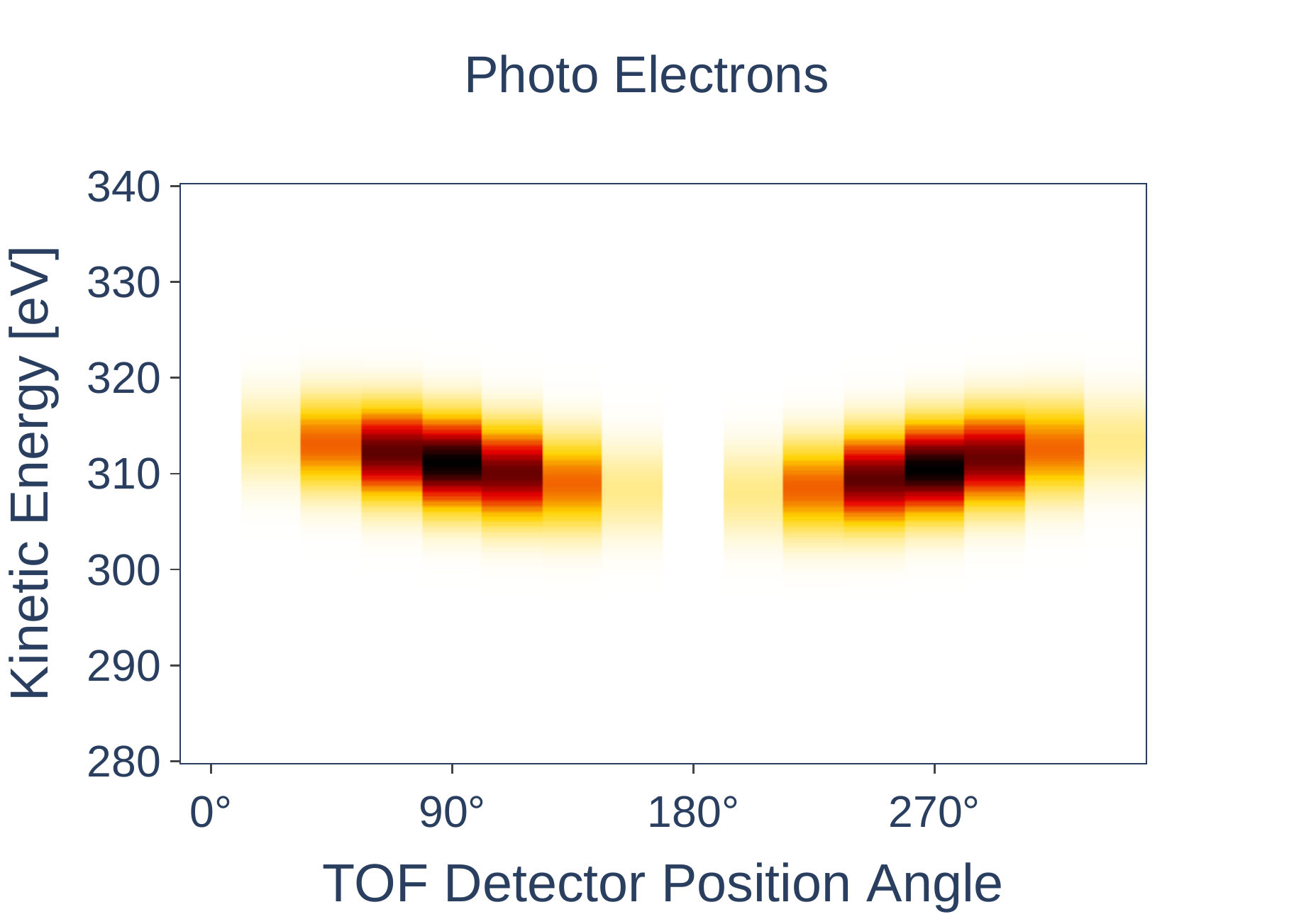}\label{fig:5b}}
    \vfill
    \subfloat[NN estimate ($7.0$ fs) of the Auger decay time target ($7.0$ fs) with a large kick ($22.5$ eV).
    ]{\includegraphics[width=0.49\textwidth]{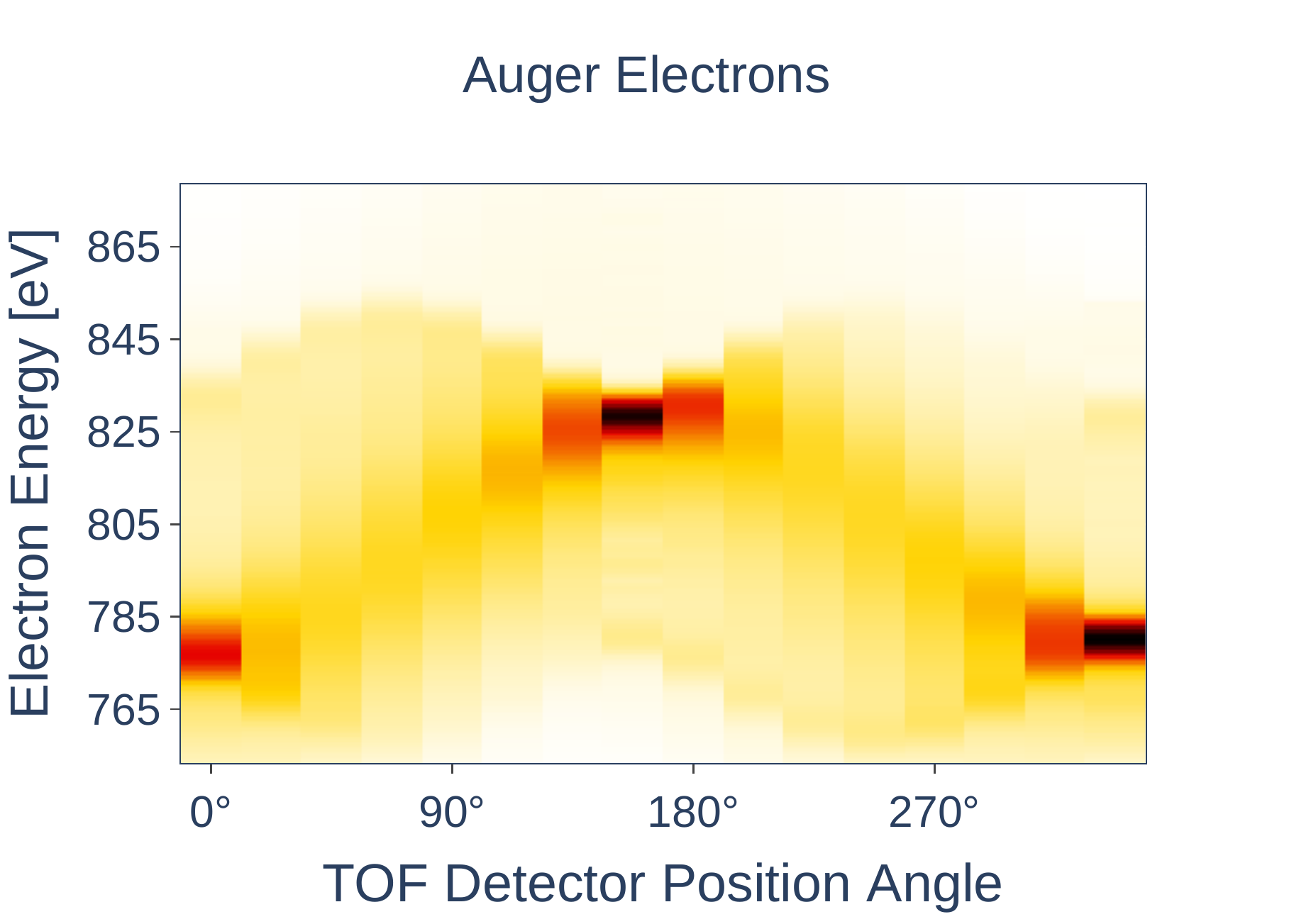}\label{fig:5c}}
    \hspace{1em}
    \subfloat[NN estimate ($7.6$ fs) of the Auger decay time target ($7.0$ fs) with a small kick ($3.0$ eV).]{\includegraphics[width=0.49\textwidth]{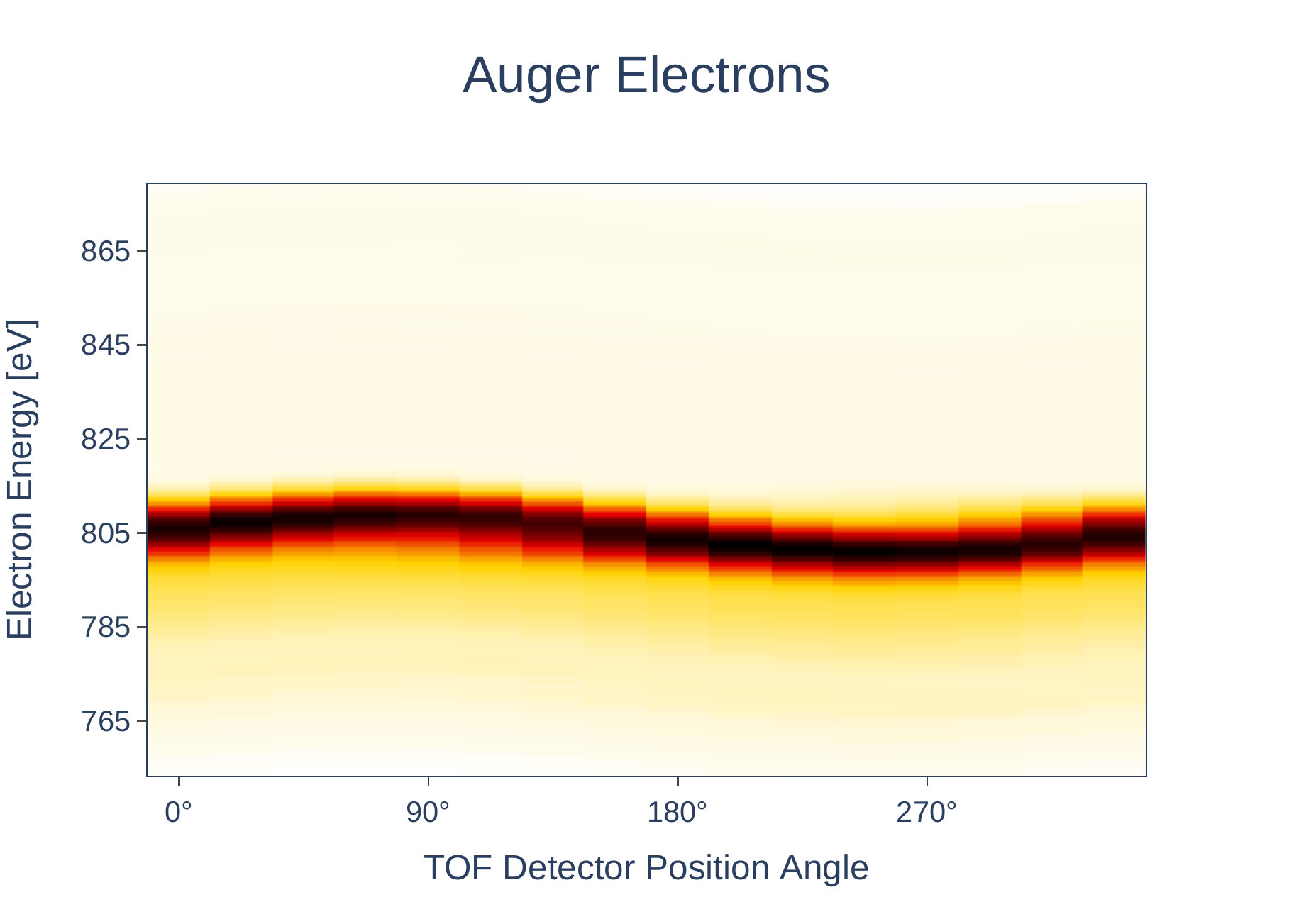}\label{fig:5d}}
    \caption{Exemplary selection of FWHM pulse duration and Auger decay time predictions with large and small kick value, respectively.}
    \label{fig:large_vs_small_kick}
    \end{figure}

\begin{figure}
    \centering
    \includegraphics[width=.6\textwidth]{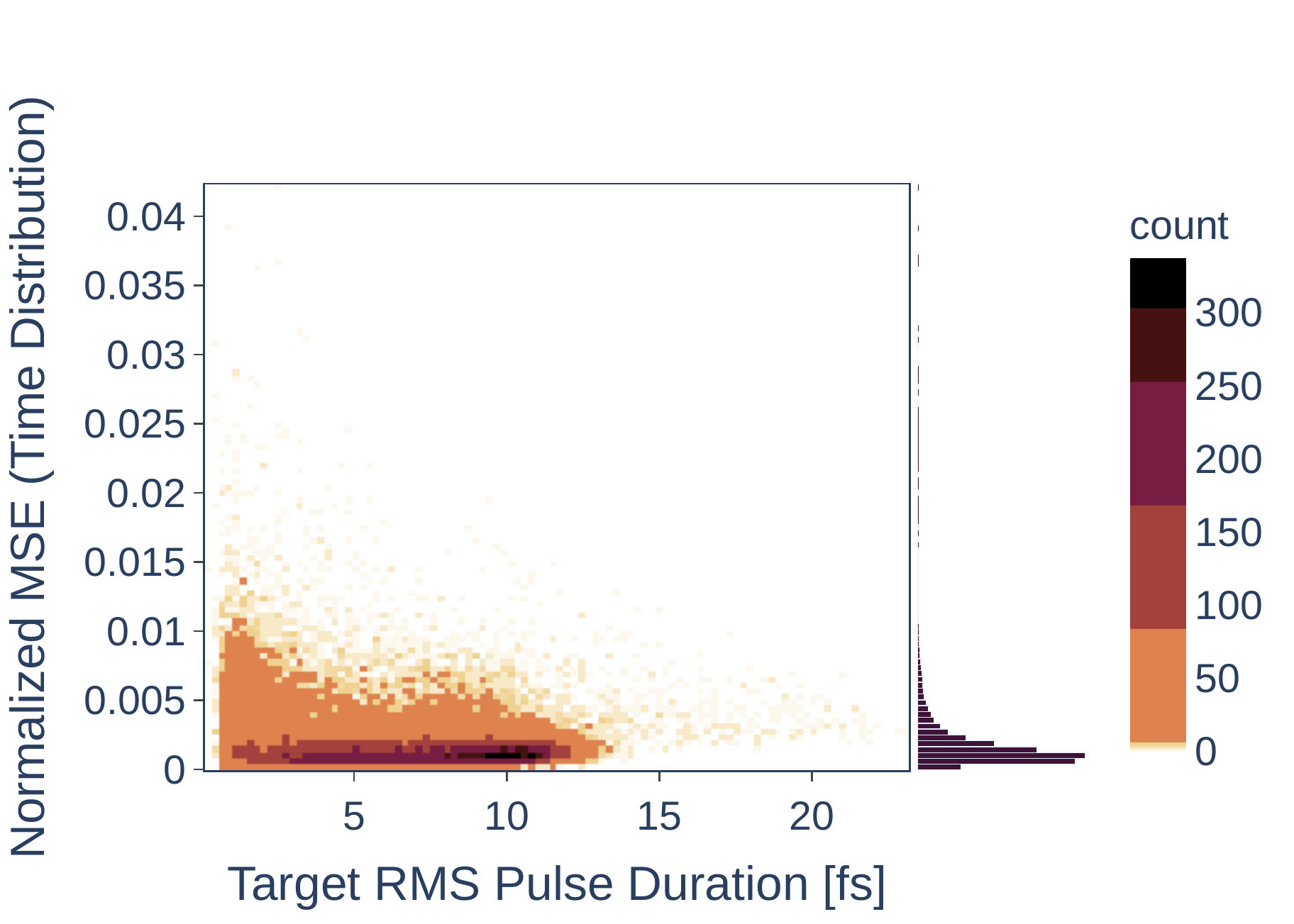}
    \caption{Mean squared error of the predicted pulse structure with regard to the true pulse structure, normalized to the respective pulse's duration, shown as a function of the true pulse structure's RMS pulse duration.}
    \label{fig:td_mse_vs_pl}
\end{figure}

\begin{figure}[!ht]
    \centering
    \subfloat[
    Estimate ($22.52$ eV) of the target kick label ($22.50$ eV). The difference is $0.02$ eV.]{\includegraphics[width=0.48\textwidth, valign=c]{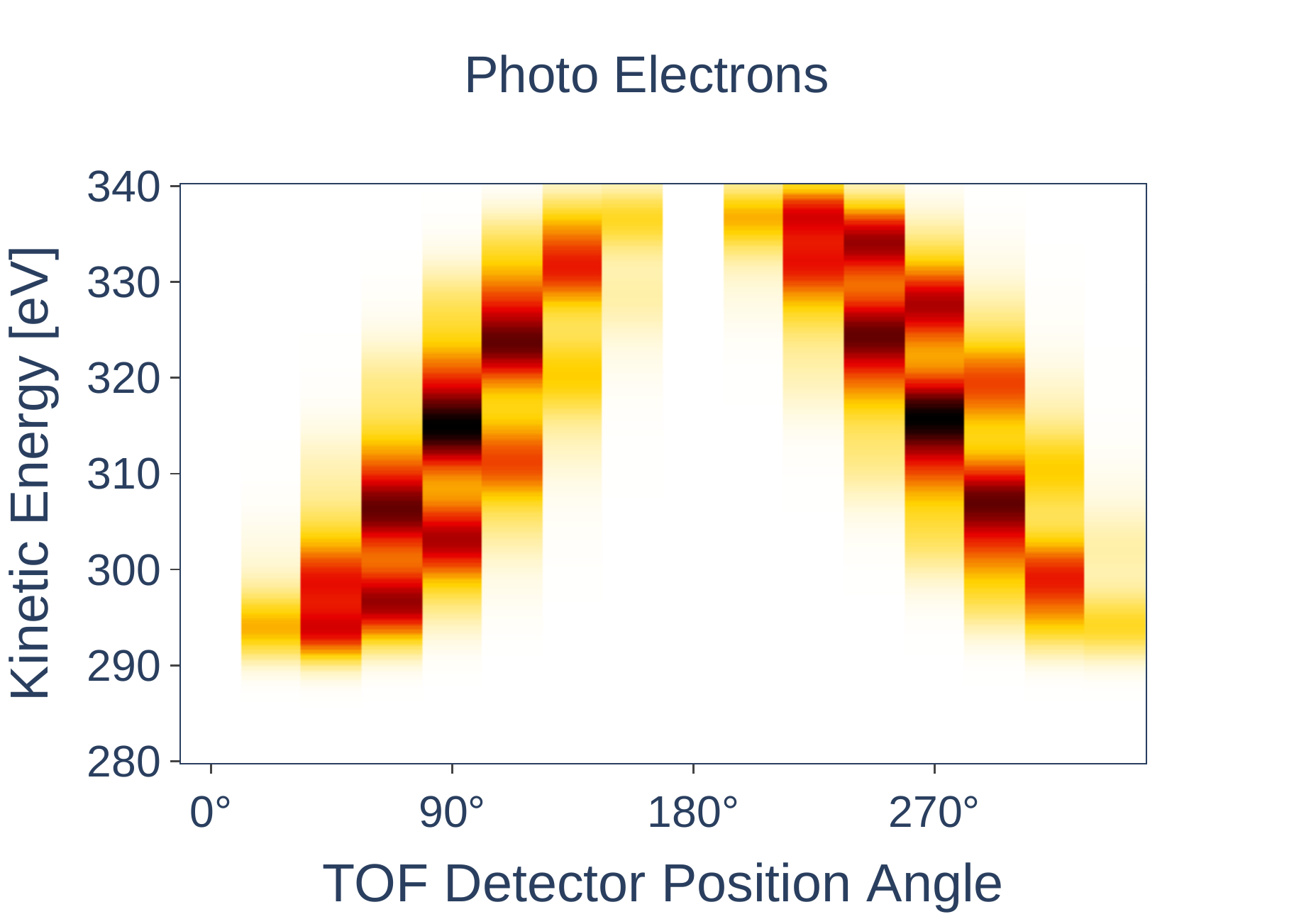}}
    \hspace{1em}
    \subfloat[
    Estimate ($22.29$ eV) of the target kick label ($22.50$ eV). The difference is $0.21$ eV.]{\includegraphics[width=0.48\textwidth,valign=c]{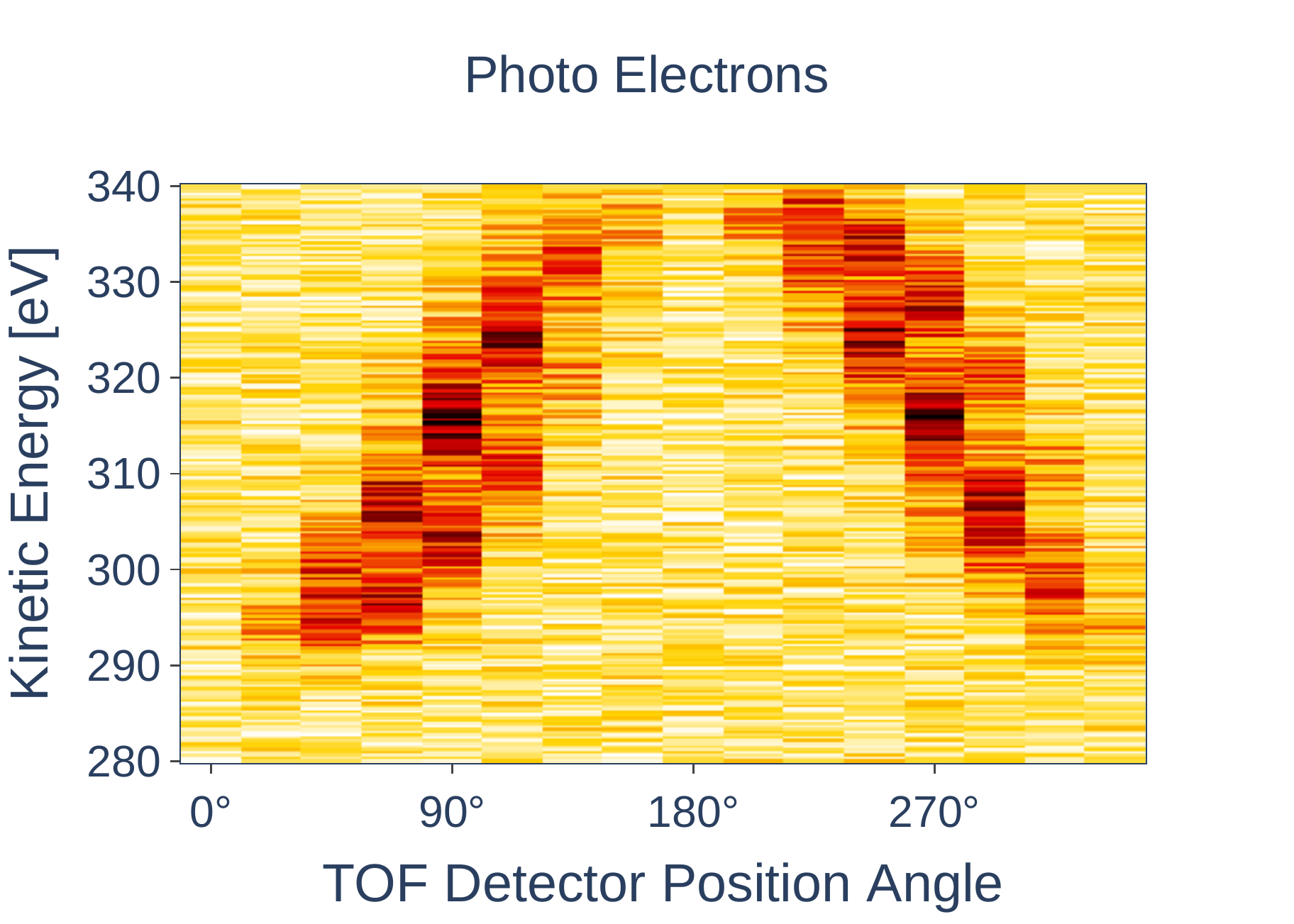}}
    \vfill
    \subfloat[
    Estimate ($4.80$ fs) of the target FWHM pulse duration label ($4.85$ fs). The difference is $0.05$ fs.]{\includegraphics[width=0.48\textwidth, valign=c]{sim_images/NN_examples/PL11_kick75_16x200_n100.h5_example_i99.pdf}}
    \hspace{1em}
    \subfloat[
    Estimate ($5.06$ fs) of the target FWHM pulse duration label ($4.85$ fs). The difference is $0.19$ fs.]{\includegraphics[width=0.48\textwidth, valign=c]{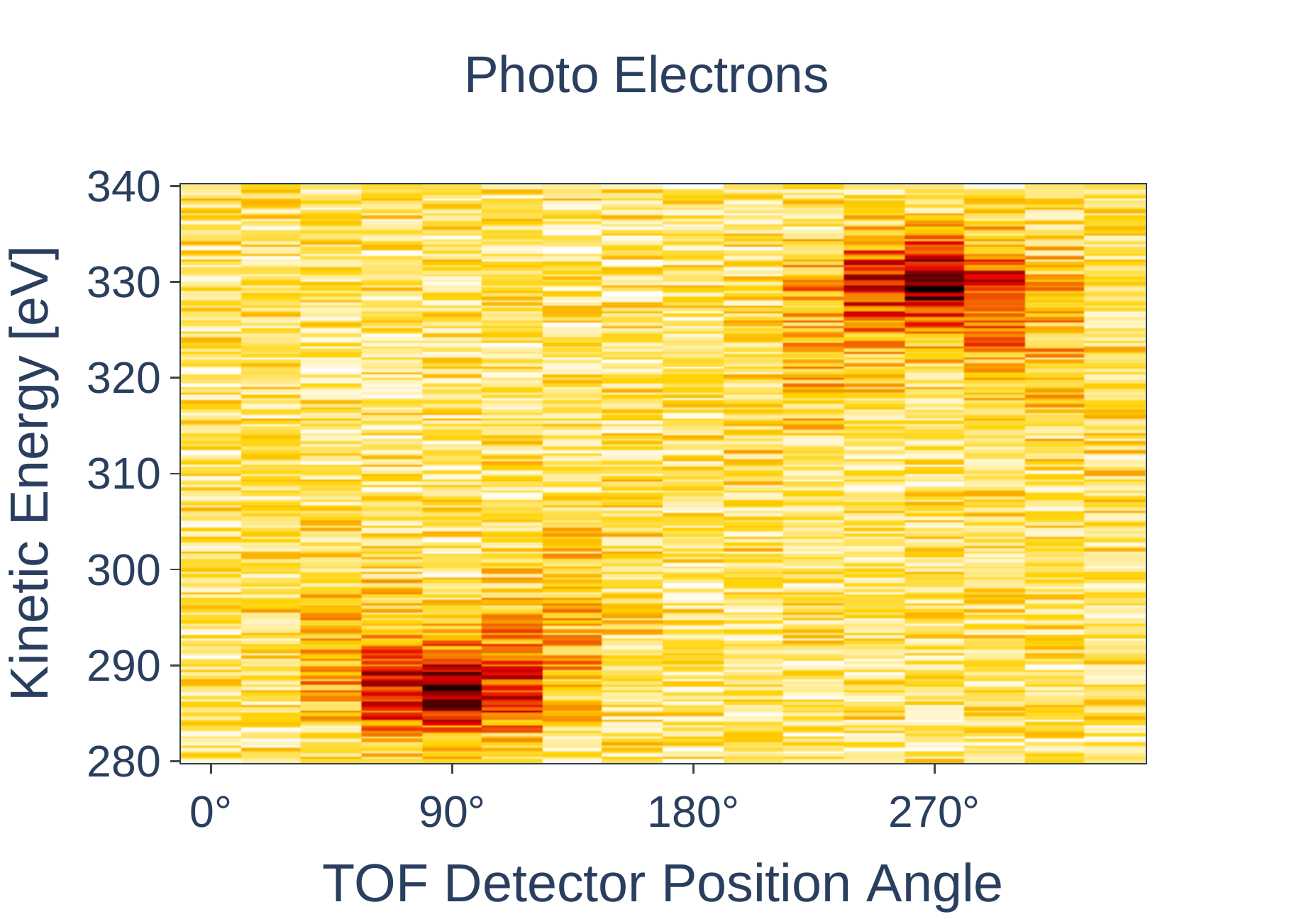}}
    \vfill
    \subfloat[
    Estimate ($7.03$ fs) of the target decay label ($7.0$ fs). The difference is $0.03$ fs.]{\includegraphics[width=0.48\textwidth, valign=c]{sim_images/NN_examples/Decay7_kick75_PL11_16x200_n100.h5_example_i0.pdf}}
    \hspace{1em}
    \subfloat[
    Estimate ($7.26$ fs) of the target decay label ($7.0$ fs). The difference is $0.26$ fs.]{\includegraphics[width=0.48\textwidth, valign=c]{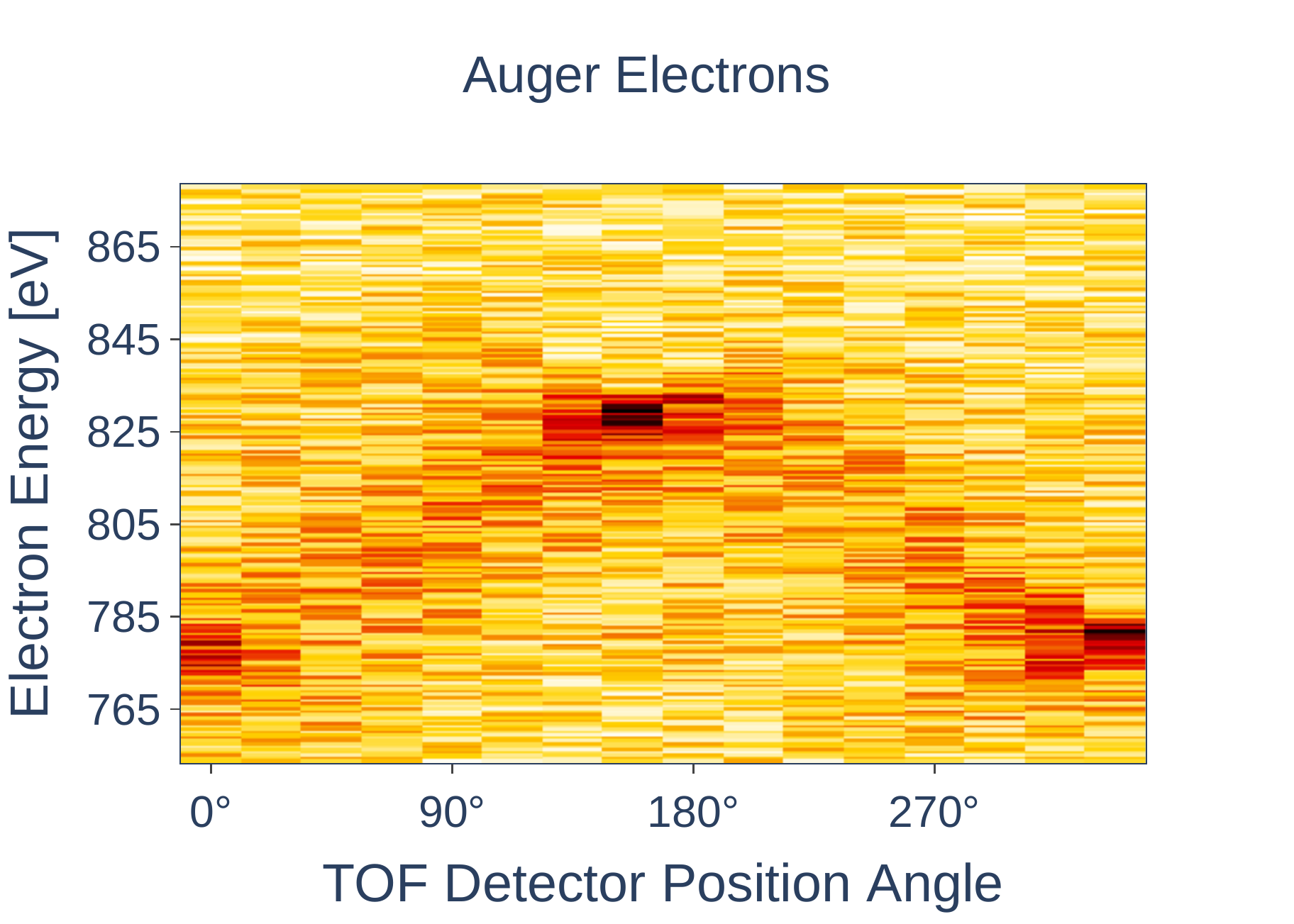}}
    \caption{The influence of noise on the respective label estimate regarding one sample. (a), (c), (e) display estimates on data without additional noise. (b), (d), (f) show estimates of data with $\pm$30 \% noise added.}
    \label{fig:noise_vs_nonoise}
    \end{figure}